\documentclass[aps,prd,showpacs,nofootinbib,floats,floatfix,preprintnumbers,groupedaddress,twocolumn]{revtex4}

\usepackage{bm}
\usepackage{latexsym}
\usepackage{dcolumn}
\usepackage{amsmath,amsfonts,amssymb}
\usepackage{graphicx,epsfig}
\usepackage{amsmath}
\usepackage{fancyhdr}
\usepackage{hyperref}
\usepackage{graphicx,epstopdf}
\hypersetup{
	colorlinks   = true, 
	urlcolor     = blue, 
	linkcolor    = blue, 
	citecolor   = red 
}

\begin{document}
\title{ Fresh look at the scalar-tensor theory of gravity in Jordan and Einstein frames from undiscussed standpoints}
\author{Krishnakanta Bhattacharya\footnote {\color{blue} krishnakanta@iitg.ernet.in}}
\author{Bibhas Ranjan Majhi\footnote {\color{blue} bibhas.majhi@iitg.ernet.in}}

\affiliation{Department of Physics, Indian Institute of Technology Guwahati, Guwahati 781039, Assam, India}

\date{\today}

\begin{abstract}
We study the scalar-tensor theory of gravity profoundly in the action level as well as in the thermodynamic level. Contrary to the usual description in the literature about the equivalence in the two conformally connected frames, this paper addresses several incomplete inferences regarding it and mentions some inequivalences which were not pointed out earlier. In the thermodynamic level, our analysis shows the two frames are equivalent. In that process, we identify the entropy, the energy and the temperature for the thermodynamic description, and we find these quantities are conformally invariant even without any prior assumption.  The same conclusion is reached from the gravitational action as well as from the Gibbons-Hawking-York boundary term, establishing the result in a more convincing manner. 
\end{abstract}

\maketitle

\section{Introduction}
It is now a well-known fact that Einstein's theory of general relativity is not considered as the complete theory to describe the Universe. Though, the theory has been played the pioneering role for decades and is highly accurate up to several precession, several incompleteness of the theory are coming out in the course of time. The theory is not consistent with the present acceleration of the Universe and fails to explain several observational data \cite{Riess:2001gk, Riess:2004nr, KNOP, Perlmutter:1998np, Tonry:2003zg, Barris:2003dq, Perlmutter:1997zf, Riess:1998cb, RIESS4}. Moreover, there are some long-standing problems in Einstein's theory of general relativity (GR), such as:
the presence of singularities in this theory (where GR fails to describe), galaxy rotation curves, difficulties in explaining the inflation model etc. Due to these drawbacks, many alternative GR theories have been introduced (for details see \cite{HEHL, WILL, Clifton:2006jh, Faraoni:2010yi, Faraoni:1998qx}). These theories are also known as modified theories of gravity which should reproduce GR in the weak-field regime or in solar length scale. However, in principle, they can differ substantially from GR in the strong curvature regime, where nonlinear effects might be dominant. The simplest picture one can think of in alternative theory of gravity case is the coupling of the gravitational field with a scalar field. Moreover, the gravitational field can also be coupled with vectors and higher-dimensional tensors in principal. But, of course, one should keep one thing in mind that those extra fields should be suppressed in the region where the GR is highly tested.

The scalar-tensor theory of gravity, pioneered by Jordan, Brans and Dicke \cite{Brans:1961sx}, is probably the most popular among those alternative GR theories.  Moreover, one of the other strong reasons to study the theory might be the fact that in the low energy limit, the bosonic string theory boils down to the Brans-Dicke theory \cite{Callan:1985ia}. Here we do our analysis for the more general scalar-tensor theory and from the scalar-tensor theory one can obtain the Brans-Dicke theory just putting the parameter $\omega(\phi)$ as a constant. It has already been proved that the Brans-Dicke theory, (special case of the scalar-tensor theory, in this case $\omega$ is not a function of the scalar field $\phi$) has been surpassed by Einstein's theory of gravity in the inflation model of the Universe  \cite{La:1989za, Laycock:1993bc, Bertolami:1999dp}. In the standard Einstein theory of general relativity, the metric tensor or the geometry is the sole quantity that describes the gravity. In this theory, another scalar field $\phi$ is introduced in addition to the spacetime metric to describe that. The scalar field in this theory automatically ends up the inflationary era without providing fine tuning of the cosmological parameters.  Moreover,  the parameter $\omega$,  which  characterizes the coupling strength of the matter and the scalar field, can be adjusted according to the demand of the situation and it is not determined by the theory a priory. It has been a conviction for long time that the theory boils down to the standard GR theory of Einstein when $ \omega\rightarrow\infty$ \cite{Weinberg} while some works refutes the statement\cite{Faraoni:1999yp, Bhadra:2002qk} for the cases when the trace of the energy momentum tensor vanishes (although recent work \cite{Pal:2016hxt} suggests that for quantized Brans-Dicke theory indeed reduces to GR when $ \omega\rightarrow\infty$, even though $T_{\mu}^{\mu}=0$). The gravitational constant $G$ is not a constant in this theory (Jordan's frame), rather the $G$ is replaced here by $G_{eff}=\phi^{-1}$. The scalar field $\phi$ is nonminimally coupled with the Ricci scalar in the Lagrangian of this frame and makes the theory to be highly nonlinear. Using the conformal  transformation of the metric tensor and rescaling the scalar field $\phi$, one can project the theory in Einstein's frame and now the Ricci scalar is not coupled with the matter in this frame rather, the scalar field is coupled with the matter field.  In the Einstein's frame the gravitational constant is a true constant.

Over the years, the scalar-tensor and the Brans-Dicke theories have been studied as two of the most popular  alternative theories of gravity (\cite{Hawking:1972qk, Campanelli:1993sm, Kang:1996rj, Koga:1998un, Faraoni:1998qx, Santiago:1999by, Dehghani:2006xt, Sheykhi:2009vc,  Capozziello:2010sc, Faraoni:2010yi, Faraoni:2016ozb}). In these works, many aspects has been studied, including ones regarding the solutions of the metric \cite{Hawking:1972qk, Campanelli:1993sm, Faraoni:2016ozb}
 , the comparative study of the two frames and their equivalence \cite{Capozziello:2010sc, Faraoni:2010yi, Koga:1998un}, the thermodynamic entities in the two frames \cite{Sheykhi:2009vc, Dehghani:2006xt, Santiago:1999by, Koga:1998un, Kang:1996rj} and others. There
are also many aspects that have been recently studied. For example, the Brans-Dicke theory has been studied with the Weyl geometry \cite{Lobo:2016izs}, where the Jordan frame is compared with the  Riemann frame in the Weyl geometry and the Einstein frame is compared with the geometrical Weyl frame. There have been many discussions in the literature about the physical (in)equivalence of the two frames and much dissonance among people in interpreting one frame as more physical than the other (for a review, see \cite{Faraoni:1998qx}). Most of the works suggest that the two frames are equivalent in the classical level, but they are not in the quantum level (one of the recent works in this context is \cite{Banerjee:2016lco}). Recently, \cite{Pandey:2016unk} finds the equivalence of the two frames in the quantum level during the absence of any external matter field and, thereby, disproves the previous conclusion mentioned in \cite{Banerjee:2016lco}. 
 Still there is no unanimous conclusion in this issue. The anomaly about the equivalence in the two frames in f(R) gravity, which is a subclass of scalar-tensor theory, has been discussed earlier \cite{Saltas:2010ga} as well as recently in \cite{New} which also addresses some earlier works in this context. It should be remembered that the mathematical equivalence of the two frames using conformal transformation does not guarantee the physical equivalence.

In this work, we shall try to explore  the equivalence of the two frames on the various aspects. Also, many things will be highlighted that are often not mentioned in the literature. Our approach is completely different compared to other works. This paper will cast light on the two frames in their dynamic level as well as in the thermodynamic point of view. To compare the two frames classically, the Lagrangian in the two frames requires detailed scrutiny to find to find their equivalence. Besides, the Gibbons-Hawking-York boundary term will be introduced in both frames, and the role of this term in the discussion of the equivalence of the two frames at the classical level will be justified. Later, a comparative study of the two gravitational actions and the same of the Einstein-Hilbert one will be provided and the special properties of the latter action (specially the holographic property of the GR) will be verified so that one can compare the theory with the usual GR case and can savour the new aspects of this theory very distinctly. This work will also discuss how the equations of motion are connected to same in the other frame. Moreover, the way of obtaining the equations of motion from the bulk part will be discussed and obliterating the contribution from the surface part will be rationalized. 

To compare the two frames in the thermodynamic level, the focus will be on obtaining the entropy in the two frames. For that, the Noether current and the potential of the two frames needs to be found as those are directly connected to the thermodynamic quantities. Then the Virasoro algebra technique will be introduced to obtain the entropy from the first principle and, thereby, the relation of entropy in the two frames can be found out. One of the other important features of the Einstein-Hilbert action is that it can be interpreted as the free energy of spacetime. This paper will also highlight that issue for the scalar-tensor theory in its two frames. From that discussion, we shall obtain the relation between the other thermodynamic quantities (the energy and the temperature) in the two frames. 
Once the relations between those quantities in the two frames gets determined from the analysis of the gravitational action, we shall verify all the conclusions from the GHY surface terms as well. 

The discussions in this paper are organized as follows: the following section will be assigned for the discussions on the equivalence of the two frames in the classical level and for the comparative study of the two frames with the usual GR case. The equivalence of the two frames at the action level and the role of the GHY term will be highlighted and, later, the total action in the two frames will be decomposed in terms of the bulk term and the total derivative surface term. After that the separation of the total Lagrangian in both the frames will be justified by obtaining the equations of motion in the two frames and by getting the holographic relation between the two terms. In the subsequent section, we shall explore the whole theory from the thermodynamic point of view. We shall obtain the Noether current and the Noether potential in the two frames and in the next part those will be used to obtain the entropy using the Virasoro algebra technique. Afterward, the relation of the energy and the temperature in the two frames will be obtained and, thereby, we shall try to interpret the action as the free energy of the spacetime. Also, the holographic relation will also be checked in the thermodynamic level as well and in the final part we discuss all the arguments in thermodynamic level form the GHY term. We shall summarize our results, and the conclusion of our analysis will be in the final section. We shall provide a brief summary at the end of each section.

\section{Comparison of the two frames at the Lagrangian level}
As mentioned earlier, there are multifarious works on the (in)equivalence of the two frames and still there are a few scopes of pointing out the areas where people can examine the (in)equivalence of the two frames. Here, the analysis starts from the action level. The theory in the action level will be studied in a great detail manner. It will be shown, while projecting the theory from one frame to the other by the conformal transformation of the metric and rescaling of the scalar field, one neglects a total derivative term to conclude that the actions in the two frames are equivalent. Therefore, the usual way of saying the mathematical equivalence of the two frames is an incomplete statement. If one incorporates the Gibbons-Hawking-York(GHY) boundary term in the analysis, then the total action(the gravitational with the GHY term) is invariant under those transformation. The holographic property will also be tested in each frame for the discussion on the equivalence of the two frames and to compare the theory with the GR after the separation of the total action into the bulk and the surface part. The results of all the analysis is mentioned in the proper places.

\subsection{Actions in the two frames from bird's eye}
The action of the gravitational field of the scalar-tensor theory in Jordan frame is given by 
\begin{eqnarray}
&&\mathcal{A}=\int d^4x\sqrt{-g}L =\int d^4x\sqrt{-g} \frac{1}{16\pi}\Big(\phi R
\nonumber
\\
&&-\frac{\omega (\phi)}{\phi}g^{ab}\nabla_a\phi \nabla_b\phi -V(\phi)\Big)~.
\label{SJ}
\end{eqnarray}
As we have mentioned earlier, in this frame the scalar field $\phi$ is nonminimally coupled with the Ricci scalar which makes the theory highly nonlinear.

With the help of the conformal transformation of the metric and the re-scaling of the scalar field $\phi$, the nonminimal coupling of the field $\phi$ with the scalar curvature in the Jordan frame goes away and one then arrives to the mathematically equivalent picture or frame which is known as the Einstein frame. The transformations are given by
\begin{align}
g_{ab}\rightarrow\tilde{g}_{ab}=\Omega^2g_{ab},\ \ \ \ \ \ \ \ \Omega=\sqrt{\phi} 
\label{GAB}
\end{align} 
and also  by the nonlinear transformation of the scalar field:
\begin{align}
 \phi\rightarrow\tilde{\phi}\,\ {\textrm{with}}\,\ d\tilde{\phi}=\sqrt{\frac{2\omega+3}{16\pi}}\frac{d\phi}{\phi}~.
\label{PHI}
\end{align}
The action of the gravitational field in the Einstein frame is given by
\begin{eqnarray}
&&\tilde{\mathcal{A}}=\int d^4x\sqrt{-\tilde{g}}\tilde{L}
\nonumber
\\
&&=\int d^4x\sqrt{-\tilde{g}}[\frac{\tilde{R}}{16\pi}-\frac{1}{2}\tilde{g}^{ab}\tilde{\nabla}_a\tilde{\phi}\tilde{\nabla}_b\tilde{\phi}-U(\tilde{\phi})]~.
\label{SE}
\end{eqnarray}
As one can see, in this frame the scalar field is not coupled with the Ricci scalar and, therefore, the gravitational part is similar to the GR case. Although we are not interested with the matter field, let us make comment that in this frame the matter field is nonminimally coupled with the scalar field $\phi$, unlike the Jordan frame.
  
    It should be mentioned here that the transformations \eqref{GAB} and \eqref{PHI} does not give the exactly same Lagrangian in the two frames mathematically, rather, they are related by
\begin{align}
16\pi\sqrt{-\tilde{g}}\tilde{L}=16\pi\sqrt{-g} L-3\sqrt{-g}\square\phi~,
\label{ACT}
\end{align}
with $U(\tilde{\phi}) = \frac{V(\phi)}{16\pi\phi^2}$.
The actions can be obtained as the same in the two frames, neglecting the total derivative term (i.e. the last term on the right-hand side of the above) as it does not affect the equation of motion. This is usually stated in the literature, but it must be noted that the actual mathematical relation is (\ref{ACT}) and to obtain the correct equations of motion, one has to fix both the variation of the field and its derivative, i.e., $\delta\phi$ and $\partial_i\delta\phi$ as zero. This is forbidden as it violates the uncertainty relation and also creates the unphysical nature of the theory. So the statement that actions (\ref{SJ}) and (\ref{SE}) are equivalent as the total derivative terms, do not contribute and, hence, can be neglected, now sounds questionable.    

  Now remember that it is the same problem that appears in the Einstein-Hilbert action and, therefore, the GHY surface term is needed with the actual gravitational action. One needs to follow the same prescription to get rid of this difficulty here also. 
Let us start with the Gibbons-Hawking-York boundary action in the Einstein frame. It is given by the standard form:
\begin{equation}
\tilde{\mathcal{A}}_{GHY}=-\frac{1}{8\pi}\int d^3x \sqrt{\tilde{h}}\tilde{K} \label{SEGHY}
\end{equation}
Here $\tilde{K}=-\tilde{\nabla}_a\tilde{N}^a=-\frac{1}{\sqrt{-\tilde{g}}}\partial_a(\sqrt{-\tilde{g}}\tilde{N}^a)$, the trace of the extrinsic curvature tensor and $\tilde{N}_a$ is the unit normal to the slice, which is spacelike or timelike, depending on the type of slice chosen; i.e. $\tilde{g}_{ab}\tilde{N}^a\tilde{N}^b=\epsilon$ where $\epsilon=+1$ spacelike while $\epsilon=-1$ for timelike normals. Under the conformal transformation (\ref{GAB}) the unit normal transforms as $\tilde{N}^a=\Omega^{-1}N^a$ such that $g_{ab}N^aN^b=\epsilon$ again. 
Then one can easily determine the relation between the $K$ and $\tilde{K}$:
\begin{equation}
\tilde{K}=\frac{1}{\Omega}K-\frac{3}{\Omega^2}N^a\partial_a\Omega  \label{K}~,
\end{equation}
where $K=-\nabla_aN^a$.
Therefore if the total action of the theory in the Einstein frame is defined as $\tilde{\mathcal{A}}_{total}=\tilde{\mathcal{A}}+\tilde{\mathcal{A}}_{GHY}$, then under the transformations (\ref{GAB}) and (\ref{PHI}) it takes the form:
\begin{eqnarray}
&&\tilde{\mathcal{A}}_{total} = \mathcal{A}-\frac{1}{8\pi}\int d^3x\sqrt{h}\Omega^2K 
\nonumber
\\
&&+\frac{3}{8\pi}\int d^3x\sqrt{h}\Omega N^a\partial_a\Omega-\frac{3}{16\pi}\int d^4x\sqrt{-g}\Box\phi~.
\end{eqnarray}
In the above (\ref{ACT}) has been used.
Now the last term on the right-hand side, by using Gauss's theorem, can be expressed as $\int d^3x\sqrt{h}N^a\partial_a\phi$, which for $\Omega=\sqrt{\phi}$ leads to the third term of the above.
Then we get $\tilde{\mathcal{A}}_{total}=\tilde{\mathcal{A}}+\tilde{\mathcal{A}}_{GHY}=\mathcal{A}+\mathcal{A}_{GHY}$ with the Gibbons-Hawking-York surface term in the Jordan frame identified as
\begin{equation}
\mathcal{A}_{GHY}=-\frac{1}{8\pi}\int  d^3x \sqrt{h}\phi K~. 
\label{SJGHY}
\end{equation}
At this point, we can say confidently that the scalar-tensor theory has an equivalent description at the action level in both the Einstein and Jordan frames that is free of any discrepancy. One should also note that if one takes the external matter field from the beginning, one finds that all the above relations are unchanged as the matter field itself is equivalent in the two frames.
\vskip 1mm
\textit{\textbf{Summary:}  While projecting the theory from one frame to the other by the conformal transformation of the metric and rescaling of the scalar field, one neglects a total derivative term to conclude that the two actions are equivalent. But, only if one incorporates the Gibbons-Hawking-York(GHY) boundary term in the analysis, then the total action(gravitational and GHY term) is invariant in the two frames.}

\subsection{Decomposition of the action as bulk term and surface term}
Let us now have a deeper inspection of the scalar-tensor theory in the two frames. In the case of the Einstein frame, the situation is similar to the GR case. It is a well-known fact that the Einstein-Hilbert Lagrangian (i.e. without the Gibbons-Hawking-York surface term) can be separated into two parts. The first one is the quadratic part $L_{quad}$, containing the $\mathcal{O}(\Gamma^2)$ terms and the second one is the total divergence term $L_{sur}$ (for details see \cite{Padmanabhan:2004fq}). Here we introduce the same discussions for the scalar-tensor theory in the Einstein and Jordan frames and we show that one can get the similar terms in these frames as well. Like the usual one, the GHY term will not be considered. So the relation (\ref{ACT}) is of importance in this section. In addition, the last total derivative term will not been thrown as we shall observe that it plays an important role in the later discussion. We start with the Lagrangian in the Einstein frame since it is similar to GR where this decomposition is known, and then we use the transformations to obtain the required decomposition in the Jordan frame.
\vskip 1mm
\noindent
{\bf Einstein frame:}
The Lagrangian of the scalar-tensor theory in the Einstein frame is given by \eqref{SE}, and it is connected with the Lagrangian of the Jordan frame by \eqref{ACT}. Now, one can show that
\begin{align}
\sqrt{-\tilde{g}}\tilde{R} = \sqrt{-\tilde{g}} \tilde{g}^{ab}(\tilde{\Gamma}^{i}_{ja}\tilde{\Gamma}^{j}_{ib}-\tilde{\Gamma}^{i}_{ab}\tilde{\Gamma}^{j}_{ij}) + \partial_{c}[\sqrt{-\tilde{g}}\tilde{V}^{c}]
\label{DEC}
\end{align}
where, $\tilde{V}^{c}=\tilde{g}^{ik}\tilde{\Gamma}^{c}_{ik}-\tilde{g}^{ck}\tilde{\Gamma}^{m}_{km}$. 
Therefore we identify the bulk term of the scalar-tensor theory in the Einstein frame as
\begin{align}
\tilde{L}_{bulk}=\frac{1}{16\pi}\tilde{g}^{ab}(\tilde{\Gamma}^{i}_{ja}\tilde{\Gamma}^{j}_{ib}-\tilde{\Gamma}^{i}_{ab}\tilde{\Gamma}^{j}_{ij}) -\frac{1}{2}\tilde{g}^{ab}\tilde{\nabla}_a\tilde{\phi}\tilde{\nabla}_b\tilde{\phi}-U(\tilde{\phi})~;
\label{QUAD}
\end{align}
whereas the surface part is given by
\begin{align}
\tilde{L}_{sur}=-\partial_c\tilde{P}^c~, 
\label{SUR}
\end{align}
where,
\begin{align}
\tilde{P}^c=-\frac{1}{16\pi}\sqrt{-\tilde{g}}\tilde{V}^c=\frac{\sqrt{-\tilde{g}}}{16\pi}(\tilde{g}^{ck}\tilde{\Gamma}^i_{ki}-\tilde{g}^{ik}\tilde{\Gamma}^c_{ik})~, \label{PCTIL}
\end{align}
such that $\sqrt{-\tilde{g}}\tilde{L}=\sqrt{-\tilde{g}}\tilde{L}_{bulk} + \tilde{L}_{sur}$.

\vskip 1mm
\noindent
{\bf Jordan frame:} Now, using the transformations (\ref{GAB}) and (\ref{PHI}), one can get the corresponding terms in the Jordan frame. Straight forward calculations give 
\begin{eqnarray}
&&\tilde{g}^{ab}(\tilde{\Gamma}^{i}_{ja}\tilde{\Gamma}^{j}_{ib}-\tilde{\Gamma}^{i}_{ab}\tilde{\Gamma}^{j}_{ij})=\Omega^2g^{ab}\sqrt{-g}[\Gamma^i_{ja}\Gamma^j_{ib}-\Gamma^{i}_{ab}\Gamma^j_{ij}]
\nonumber
\\
&&-2\Omega^2g^{ab}\sqrt{-g}\Gamma^i_{ab}(\partial_i\ln\Omega)
\nonumber
\\
&&+6\sqrt{-g}\Omega^2(\partial^i\ln\Omega)(\partial_i\ln\Omega)
\nonumber
\\
&&+2\sqrt{-g}\Omega^2\Gamma^i_{ij}(\partial^j\ln\Omega)~;
\end{eqnarray}
and
\begin{align}
\sqrt{-\tilde{g}}\tilde{V}^{c}=\Omega^2\sqrt{-g}(g^{ik}\Gamma^c_{ik}-g^{ck}\Gamma^m_{km})-6\Omega^2\sqrt{-g}\partial^c(\ln\Omega)~. \label{VC}
\end{align}
Therefore, one obtains
\begin{eqnarray}
&&\sqrt{-\tilde{g}}\tilde{L}=(1/16\pi)\Big[\Omega^2g^{ab}\sqrt{-g}[\Gamma^i_{ja}\Gamma^j_{ib}-\Gamma^{i}_{ab}\Gamma^j_{ij}]
\nonumber
\\
&&-2\Omega^2g^{ab}\sqrt{-g}\Gamma^i_{ab}(\partial_i\ln\Omega)
+2\sqrt{-g}\Omega^2\Gamma^i_{ij}(\partial^j\ln\Omega)\Big]
\nonumber
\\
&&-\frac{4}{16\pi}\sqrt{-g}\omega\Omega^2(\partial_i\ln\Omega)(\partial^i\ln\Omega)-\frac{V(\phi)}{16\pi\phi^2}
\nonumber
\\
&&+(1/16\pi)\Big[\partial_c[\Omega^2\sqrt{-g}(g^{ik}\Gamma^c_{ik}-g^{ck}\Gamma^m_{km})]
\nonumber
\\
&&-\partial_c[6(\Omega^2\sqrt{-g}\partial^c(\ln\Omega)]\Big]
 \label{PART1}
\end{eqnarray}
where one needs to use that $U(\tilde{\phi})=V/16\pi\phi^2$.
Using (\ref{ACT}) one can get the action in the Jordan frame expressed in bulk and surface terms. These are given by
\begin{eqnarray}
&&L_{bulk}=(1/16\pi)\Big[\Omega^2g^{ab}[\Gamma^i_{ja}\Gamma^j_{ib}-\Gamma^{i}_{ab}\Gamma^j_{ij}]
\nonumber
\\
&&-2\Omega^2g^{ab}\Gamma^i_{ab}(\partial_i\ln\Omega)
+2\Omega^2\Gamma^i_{ij}(\partial^j\ln\Omega)\Big]
\nonumber
\\
&&-\frac{4}{16\pi}\omega\Omega^2(\partial_i\ln\Omega)(\partial^i\ln\Omega)-\frac{V(\phi)}{16\pi\phi^2}
\label{bulk}
\end{eqnarray}
and
\begin{eqnarray}
&& L_{sur}=\frac{1}{16\pi}\partial_c(\Omega^2\sqrt{-g}V^c)
\nonumber
\\
&&=(1/16\pi)\partial_c[\Omega^2\sqrt{-g}(g^{ik}\Gamma^c_{ik}-g^{ck}\Gamma^m_{km})]
\label{3}
\end{eqnarray}
where one can identify $V^c=g^{ik}\Gamma^c_{ik}-g^{ck}\Gamma^m_{km}$,
such that $\sqrt{-g}L = \sqrt{-g}L_{bulk}+L_{sur}$.

  Now to test that the division of the Lagrangian in the bulk and surface terms in the Jordan frame is perfectly alright, we show in the next section that the correct equations of motions can be obtained solely from the bulk term.

\subsection{Equation of motion from the bulk term} \label{EOM}
    The division of surface and bulk terms in Einstein frame is obvious as the action is similar to usual Einstein-Hilbert action for which such terms are already well known. For confirmation of the correct identification of the two terms in the Jordan frame, we find the equations of motion from the bulk part and show that these are the correct ones, thereby proving the validity of the divisions (\ref{bulk}) and (\ref{3}).
\vskip 1mm    
\noindent
{\bf Einstein frame:}
We have $\sqrt{-\tilde{g}}\tilde{L}_{bulk}=\sqrt{-\tilde{g}}\tilde{L}-\tilde{L}_{sur}$ and, hence, the arbitrary variation of the bulk term is given by
\begin{equation}
\delta(\sqrt{-\tilde{g}}\tilde{L}_{bulk})= \delta(\sqrt{-\tilde{g}}\tilde{L}) - \delta(\tilde{L}_{sur})~, 
\end{equation}
where 
\begin{eqnarray}
&&\delta(\sqrt{-\tilde{g}}\tilde{L})=
\sqrt{-\tilde{g}}[\frac{\tilde{G}_{ab}}{16\pi}-\frac{1}{2}\tilde{\nabla}_a\tilde{\phi}\tilde{\nabla}_b\tilde{\phi}+\frac{1}{4}\tilde{g}_{ab}\tilde{\nabla}^i\tilde{\phi}\tilde{\nabla}_i\tilde{\phi}
\nonumber
\\
&&+\frac{1}{2}\tilde{g}_{ab}U(\tilde{\phi})]\delta\tilde{g}^{ab}
+\sqrt{-\tilde{g}}[\tilde{\nabla}_a\tilde{\nabla}^a\tilde{\phi}-\frac{dU}{d\tilde{\phi}}]\delta\tilde{\phi}
\nonumber
\\
&&+\sqrt{-\tilde{g}}\tilde{\nabla}_a[\frac{\delta\tilde{v}^a}{16\pi}-(\tilde{\nabla}^a\tilde{\phi})\delta\tilde{\phi}]
\label{1}
\end{eqnarray}
and 
\begin{eqnarray}
&&\delta(\tilde{L}_{sur}) = (1/16\pi)\partial_a\Big(\delta(\sqrt{-\tilde{g}}\tilde{V}^a)\Big)
\nonumber
\\
&&=\frac{1}{16\pi}\partial_a\Big[-\frac{1}{2}\sqrt{-\tilde{g}}\tilde{g}_{ik}\delta \tilde{g}^{ik}\tilde{V}^a+\sqrt{-\tilde{g}} \delta \tilde{v}^a 
\nonumber 
\\
&&+\sqrt{-\tilde{g}}(\delta \tilde{g}^{ik}\tilde{\Gamma}^a_{ik}-\delta\tilde{g}^{ak}\tilde{\Gamma}^m_{km})\Big]~.
\label{2}
\end{eqnarray}
Here $\delta\tilde{v}^a=2\tilde{P}^{ibad}\tilde{\nabla}_b\delta \tilde{g}_{id}$ and $\tilde{P}^{iabd}=\frac{\partial \tilde{R}}{\partial \tilde{R}_{iabd}}=\frac{1}{2}[\tilde{g}^{ia}\tilde{g}^{bd}-\tilde{g}^{id}\tilde{g}^{ab}]$.
Usually in the literature the equations of motion for metric and $\tilde{\phi}$ are obtained from (\ref{1}). But this is not conceptually correct as here one needs to fix both the metric and its first derivative. Interestingly this does not happen for bulk term as the problematic term $\delta\tilde{v}^a$ cancels out as it appears in both the $\tilde{L}$ and surface terms. Now imposition of the condition that the fields $\tilde{g}_{ab}$ and $\tilde{\phi}$ are fixed at the boundary, leads to the correct equations of motion:
\begin{align}
\frac{\tilde{G}_{ab}}{16\pi}-\frac{1}{2}\tilde{\nabla}_a\tilde{\phi}\tilde{\nabla}_b\tilde{\phi}+\frac{1}{4}\tilde{g}_{ab}\tilde{\nabla}^i\tilde{\phi}\tilde{\nabla}_i\tilde{\phi}+\frac{1}{2}\tilde{g}_{ab}U(\tilde{\phi})=0 \label{EOM1}
\end{align}
and 
\begin{align}
\tilde{\nabla}_a\tilde{\nabla}^a\tilde{\phi}=\frac{dU}{d\tilde{\phi}} \label{EOM2}
\end{align}

\vskip 1mm
\noindent
{\bf Jordan frame:} In the case of the theory in the Jordan frame, one obtains the same result. Here
\begin{eqnarray}
&&\delta(\sqrt{-g}L_{bulk})=\delta(\sqrt{-g}L)-\delta L_{sur}
\nonumber
\\ 
&&=\frac{\sqrt{-g}}{16\pi}(\phi G_{ab}+\frac{\omega}{2\phi}\nabla_i\phi\nabla^i\phi g_{ab}
\nonumber
\\
&&-\frac{\omega}{\phi}\nabla_a\phi\nabla_b\phi +\frac{V}{2}g_{ab}-\nabla_a\nabla_b\phi+\nabla_i\nabla^i\phi g_{ab})\delta g^{ab}
\nonumber
\\
&&+\frac{\sqrt{-g}}{16\pi}(R-\frac{1}{\phi}\frac{d\omega}{d\phi}\nabla_i\phi\nabla^i\phi +2\nabla_a(\frac{\omega}{\phi}\nabla^a\phi)-\frac{dv}{d\phi}
\nonumber
\\
&&+\frac{\omega}{\phi^2}\nabla_a\phi \nabla^a\phi)\delta\phi 
+\frac{\sqrt{-g}}{16\pi}\nabla_a[-2g^{ab}\frac{\omega}{\phi}(\nabla_b\phi) \delta\phi +\phi \delta v^a
\nonumber
\\
&&-2(\nabla_b\phi)p^{iabd}\delta g_{id}])-(1/16\pi)\delta\partial_c(\phi\sqrt{-g}V^c) \label{4}
\end{eqnarray} 
where $\delta v^a=2p^{ibad}\nabla_b \delta g_{id}$ and $p^{iabd}=\frac{\partial R}{\partial R_{iabd}}= \frac{1}{2}[g^{ia}g^{bd}-g^{id}g^{ab}]$. The last term in \eqref{4} comes from the variation of the surface part and the rest of the previous terms come from the variation of the total Lagrangian.
Like the earlier one, we can apply the same arguments here to obtain the equations of motion from only the bulk term of  the total Lagrangian. The problematic total derivative terms are almost the same as earlier, and they are overall multiplied by $\phi$ inside the derivative. The equations of motions are
\begin{eqnarray}
&&\phi G_{ab}+\frac{\omega}{2\phi}\nabla_i\phi\nabla^i\phi g_{ab}-\frac{\omega}{\phi}\nabla_a\phi\nabla_b\phi +\frac{V}{2}g_{ab}-\nabla_a\nabla_b\phi
\nonumber
\\
&&+\nabla_i\nabla^i\phi g_{ab}=0 \label{EOM3}
\end{eqnarray}
and, using the above equation of $g^{ab}$ the coefficient of $\delta\phi$ gives the equation of motion of the field $\phi$ as
\begin{align}
\square\phi=\frac{1}{2\omega+3}(\phi\frac{dV}{d\phi}-\frac{d\omega}{d\phi}\nabla^i\phi\nabla_i\phi-2V) \label{EOM4}
\end{align}
So, one can conclude that the equations of motion can be obtained only from the bulk part of the Lagrangian. Note that here in the calculation we have taken the contribution of the last term of \eqref{ACT} to identify the bulk and the surface parts of the actions and, thereby, obtain the correct equations of motion by a physically consistent prescription. Thus, the dropping of that surface term is unjustified.

It should be mentioned here that the equation of motions of $\tilde{g}^{ab}$ and $g^{ab}$(or $\tilde{\phi}$ and $\phi$) in the two frames are connected with each other by the transformation of the metrics given by \eqref{GAB} and \eqref{PHI}. In other word, if the equation of motion of say, the metric tensor (or the scalar field) is given in any frame, the mentioned transformation relations give the same in the another frame. We shall use this result in our later discussions. Moreover, here one can conclude that in the dynamic level the two frames are equivalent.
\vskip 1mm
\textit{\textbf{Summary:} The identification of the surface term and the bulk term for both the frames are consistent with the analysis made in this section. In both the frames, we have obtained the equations of motions only from the bulk part of the total action, which is the correct method as mentioned. Not, only the total action, we have shown that the equations of motions are invariant as well under the transformations (\eqref{GAB} and \eqref{PHI}) in the two frames.}

\subsection{Connection of the surface and the bulk terms of the action: The holographic relation} \label{HOL}
One of the most striking feature of the Einstein-Hilbert action in GR is that, the surface and the bulk part of the Lagrangian are connected to each other (for details see the original works \cite{Padmanabhan:2002jr,Padmanabhan:2004fq} for GR and \cite{Mukhopadhyay:2006vu} for Lanczos-Lovelock gravity). The basic idea is as follows:
 It can be shown that for any given Lagrangians, say $L_1$($q$, $\partial q$) and $L_2$($q$, $\partial q$, $\partial^2q$) which are related as
\begin{align}
L_2(q, \partial q, \partial^2q)=L_1-\partial(q\frac{\partial L_1}{\partial(\partial q)}), \label{REL}
\end{align}  
 the same equation of motion is produced when one exteremizes the actions. For the former case one has to fix $q$ and for the latter case one has to fix $\frac{\partial L_1}{\partial \dot{q}}$ (which is the canonical momentum) on the boundary. Remarkably, the last total derivative of \eqref{REL} term matches with the surface part of the Lagrangian in GR when $L_1$ is the bulk part ( which is a function of $g^{ab} $ and $\partial{g}^{ab}$ ) and $L_2$ is the total Lagrangian (which is a function of $g^{ab} $,$\partial{g}^{ab}$ and $\partial^2{g}^{ab}$). This relation in literature is known as the holographic relation\cite{Padmanabhan:2004fq}.

Here we check whether we get the same result for the scalar-tensor theory in the two frames as well.
 
\vskip 1mm

\noindent

{\bf Einstein frame:}
The action in the Einstein frame is quite similar to the Einstein-Hilbert action, and the Lagrangian in this frame is the function of the metric tensor $\tilde{g}^{ab}$ and its first- and  second-order derivatives,  but it depends only on $\tilde{\phi}$ and its first-order derivative. In spite of that, differentiation of $\sqrt{-\tilde{g}}\tilde{L}_{bulk}$ with respect to only the $\partial_c\tilde{g}_{ab}$ gives the desired result, and we do not need any term containing the differentiation of the bulk  Lagrangian with respect to $\partial_i\tilde{\phi}$. This is because the total Lagrangian is not the function of the second-order derivative of $\tilde{\phi}$. From \eqref{REL}, one can say that $q$ cannot be $\tilde{\phi}$. Otherwise, the total Lagrangian will be the function of the second-order derivative of $\tilde{\phi}$, which we cannot allow here.One can check that the holographic property is maintained here in this frame and, a straightforward calculation gives the relation between the surface and the bulk part as
\begin{align}
\sqrt{-\tilde{g}}\tilde{L}=\sqrt{-\tilde{g}}\tilde{L}_{bulk}-\partial_c[\frac{\partial\sqrt{-\tilde{g}}\tilde{L}_{bulk}}{\partial\tilde{g}_{ij,c}}\tilde{g}_{ij}]. \label{LEINTIL}
\end{align}
Here the last term is the surface term. So, it is obvious from the above that this can be obtained from the bulk part of the Lagrangian.
Thus, as is done for the Einstein-Hilbert action \cite{Padmanabhan:2004fq}, one can get the same relations for the scalar-tensor theory in the Einstein frame as well. Therefore, like the Einstein-Hilbert case, one can correlate the bulk part of the Lagrangian $\tilde{L}_{bulk}$ with $L_1$($q$, $\partial q$) and the total gravitational Lagrangian with $L_2$($q$, $\partial q$, $\partial^2q$) and one can conclude that in this frame the gravity is ``holographic". 

\vskip 1mm
\noindent
{\bf Jordan frame:}
Let us now check whether the same conclusion can be drawn for the theory in Jordan frame as well. We have already separated the Lagrangian in the Jordan frame in bulk part and the total derivative surface part \eqref{bulk} and \eqref{3}. Here also, the total Lagrangian is the function of $g^{ab}$ and it first- and second-order derivative, but it depends solely on $\phi$ and its first-order derivative. Thus, here also we can apply the same arguments and we cannot allow $q$ to be $\phi$ in \eqref{REL} to obtain the holographic relation. We perform the same steps as done in Einstein's frame and find
\begin{eqnarray}
&&\frac{\partial\sqrt{-g}L_{bulk}}{\partial g_{ab,c}}g_{ab}=-\frac{1}{16\pi}\Omega^2\sqrt{-g}V^c
\nonumber
\\
&&+\frac{6}{16\pi}(\Omega^2\sqrt{-g}\partial^c(\ln\Omega)) \label{EQU}
\end{eqnarray}
Note that the right-hand side of the above equation is the surface part plus an extra term. So, unlike in the Einstein frame, one cannot draw the same conclusion for the theory in the Jordan frame. This is significant inequivalence of the two frame at the classical level even without considering any matter field.
\vskip 1mm
\textit{\textbf{Summary:} We get in-equivalence of the two frames in the classical level. In the Einstein frame, the holographic property is valid while in the Jordan frame the property is not maintained.}
\vskip 1mm
So far, we have made our analysis at the action level. Let us now compare the two frames in the thermodynamic level.

\section{Entropy from Noether Current and the Noether Potential in the Two Frames}
Bekenstein-Hawking formula of entropy \cite{Bekenstein:1973ur} says that for a black hole in GR, the entropy is proportional to its horizon area. This was one of the earlier works that shows the connection the spacetime geometry with the gravitational thermodynamics and this connection seems more convincing with the passage of time. In our earlier work \cite{Bhattacharya:2016kbm}, we have shown that the thermodynamic structure is still maintained for the realistic time dependent black holes as well. However, it was believed that there should be a much more general expression of entropy, of which Bekenstein-Hawking formula is just first-order approximation, and the general formula should be valid for any arbitrary dimension in any theory. For the generalization of the expression of entropy, an operative definition of the black hole entropy was essential. Meanwhile, Wald obtained a direct relation of the entropy with the Noether potential \cite{Wald:1993nt} and the entropy is a conserved charge when the Lagrangian is considered to be diffeomorphism invariant. But, in this method one has to put the factor of the surface gravity by hand. Later, for the Brans-Dicke theory (not the scalar-tensor one), Kang \cite{Kang:1996rj} formulated the entropy being proportional to the area and the scalar field $\phi$ from the argument of the nondecreasing surface area. It has been shown that due to the presence of the scalar field $\phi$ in Brans-Dicke theory the surface area of a black hole becomes oscillatory during the dynamical evolution, violating the area law(that the horizon area of a black hole is an increasing function of time) which is valid in Einstein's gravity. Kang's prescription was to take the entropy of a black hole in this theory as $S_{BH}=\frac{1}{4}\int d^2x \sqrt{h}=\frac{\phi A}{4}$.

This section discusses about the procedure to get the Noether current and the Noether potential in the two frames. After that, using those quantities, entropy will be obtained from the first principle. 

\subsection{Noether current and charge} 
{\bf Jordan frame:}
The variation of the total Lagrangian in Jordan frame is shown earlier in \eqref{4}. The on-shell (i.e. using the equations of motion) expression of the variation of the total Lagrangian in the Jordan frame appear as
\begin{align}
&&\delta(\sqrt{-gL})=\frac{\sqrt{-g}}{16\pi}\nabla_a[-2g^{ab}\frac{\omega}{\phi}(\nabla_b\phi) \delta\phi +\phi \delta v^a
\nonumber
\\
&&-2(\nabla_b\phi)p^{iabd}\delta g_{id}]
\end{align}
Under the diffeomorphism $x^a\rightarrow x^a+\xi^a$, the $\delta$ is taken to be Lie derivative. Therefore $\pounds_{\xi}\phi=\xi^a\nabla_a\phi$, $\delta g_{ab}=\nabla_a\xi_b+\nabla_b\xi_a$ and $\pounds_{\xi} g^{ab}=-(\nabla^a\xi^b+\nabla^b\xi^a)$. Since $L$ is a scalar, $\delta(\sqrt{-g}L)\equiv\pounds_{\xi}(\sqrt{-g}L)=\sqrt{-g}\nabla_a(L\xi^a)$. In that case one gets the relation $\nabla_aJ^a=0$, where $J^a$ can be identified as the conserved Noether current, given by\\
\begin{eqnarray}
&&J^a=\frac{1}{16\pi}[16\pi L\xi^a+2g^{ab}\frac{\omega}{\phi}(\nabla_b\phi) \pounds_\xi\phi -\phi\pounds_\xi v^a
\nonumber
\\
&&+2(\nabla_b\phi)p^{iabd}\pounds_\xi g_{id}]
\nonumber
\\
&&=\frac{1}{16\pi}[(\phi R-\frac{\omega (\phi)}{\phi}g^{ab}\nabla_a\phi \nabla_b\phi -V(\phi))\xi^a 
\nonumber
\\
&&+2\frac{\omega}{\phi}(\nabla^a\phi)\xi^b (\nabla_b\phi)
-\phi\pounds_\xi v^a
\nonumber
\\
&&+2(\nabla_b\phi)p^{iabd}\pounds_{\xi}g_{id}]. \label{6}
\end{eqnarray}
Using the equation of motion for the field $g^{ab}$ \eqref{EOM3} and also using the explicit form of $\pounds_\xi v^a$ and $p^{iabd}$ one gets the expression of the on-shell Noether current as (for the calculations in detail, see Appendix \ref{APPEN1})
\begin{equation}
J^a=\frac{1}{16\pi}\nabla_b[\phi(\nabla^a\xi^b-\nabla^b\xi^a)+2\xi^a(\nabla^b\phi)-2\xi^b(\nabla^a\phi)]                    \label{JAJ}
\end{equation}
Therefore, the antisymmetric Noether potential $J^{ab}$, defined as $\nabla_bJ^{ab}=J^a$, turns out to be
\begin{align}
J^{ab}=\frac{1}{16\pi}[\phi(\nabla^a\xi^b-\nabla^b\xi^a)+2\xi^a(\nabla^b\phi)-2\xi^b(\nabla^a\phi)] 
\label{NOPJ}
\end{align}

\vskip 1mm
\noindent
{\bf Einstein Frame:}
We do the same for the Einstein frame.
The variation of the total action in the Einstein frame is shown earlier in \eqref{1}, on-shell expression of which gives the surface term as
\begin{equation}
\delta(\sqrt{-\tilde{g}}\tilde{L})=\tilde{\nabla}_a[\frac{\delta\tilde{v}^a}{16\pi}-(\tilde{\nabla}^a\tilde{\phi})\delta\tilde{\phi}] \label{5}
\end{equation}
 In the case of the diffeomorphism symmetry, when $\tilde{x}^a\rightarrow \tilde{x}^a+\tilde{\xi}^a $, the $\delta$ is taken to be the Lie derivative in this frame and it leads to the relation as $\tilde{\nabla}_a\tilde{J}^a=0$, where one can identify $\tilde{J}^a$ as the conserved Noether current for the diffeomorphism symmetry in the Einstein frame given by
\begin{equation}
\tilde{J}^a=\tilde{L}\tilde{\xi}^a-\frac{\pounds_{\xi}\tilde{v}^a}{16\pi}+(\tilde{\nabla}^a\tilde{\phi})\pounds_{\xi}\tilde{\phi} \label{JE}
\end{equation}
Now, using the equation of motion for the field $\tilde{g}^{ab}$ \eqref{EOM1}, one can obtain the on-shell Noether current as (for details see Appendix \ref{APPEN1})
\begin{equation}
\tilde{J}^a=\frac{1}{16\pi}\tilde{\nabla}_b[\tilde{\nabla}^a\tilde{\xi}^b-\tilde{\nabla}^b\tilde{\xi}^a] \label{JE1}
\end{equation}
Again, one can identify the Noether potential $\tilde{J}^{ab}$ in the Einstein frame as
\begin{equation}
\tilde{J}^{ab}=\frac{1}{16\pi}[\tilde{\nabla}^a\tilde{\xi}^b-\tilde{\nabla}^b\tilde{\xi}^a] \label{JABVE}
\end{equation}
In the following section we use these expressions of the Noether currents and the Noether potentials of the two frames to obtain the entropy using the Virasoro algebra technique. One can ask why we do not follow Wald's method \cite{Wald:1993nt} to get the entropy, where the calculations are much more simplistic. The answer is: Using Wald's prescription, one can obtain the entropy for the GR cases where it is given by a quarter of the horizon surface area, and for that, one has to incorporate a factor of surface gravity by hand with the conserved Noether current to obtain it. However, in the scalar-tensor theory, we do not know the expression of the entropy, and, therefore, we cannot predict whether the inclusion of the factor of the surface gravity with the conserved Noether current gives entropy in this theory. Therefore, we want to get entropy from the first principle using the Virasoro algebra technique. This method was originally introduced by Brown and Henneaux \cite{Brown:1986nw} and was further developed by Carlip \cite{Carlip:1999cy} in the near horizon symmetry. We follow this technique to obtain the entropy for the scalar-tensor theory of gravity.

\vskip 1mm


\subsection{Virasoro Algebra and the Entropy}
In this section, we need some important relations which is given in Appendix A of \cite{Carlip:1999cy} and derived in Appendix B of \cite{Majhi:2011ws}. All the following calculations in this section are done in the near horizon limit.

In this approach, we need the definition of charge and the bracket among the charges. These are derived in \cite{Majhi:2011ws} as follows
\begin{align}
Q[\xi]=\frac{1}{2}\int d\Sigma_{ab}\sqrt{h}J^{ab} \label{CHARGE}
\end{align}
and
\begin{eqnarray}
&&[Q_1,Q_2]=\int\sqrt{h}d\Sigma_{ab}[\xi_2^aJ^b[\xi_1]-\xi_1^aJ^b[\xi_2]]
\nonumber
\\
&&=\int\sqrt{h}d\Sigma_{ab}[\xi_2^aJ^b_1-\xi_1^aJ^b_2] \label{Q1Q2}
\end{eqnarray} 
Here, $h$ is the determinant of the induced 2-metric and $d\Sigma_{ab}$ is the two-dimensional surface element. The calculation of this charge and its bracket is one of the key aspects of this process. For different theories this two quantities are calculated by one of the authors \cite{Majhi:2011ws, Majhi:2012tf, Majhi:2012nq, Majhi:2013lba, Majhi:2014lka, Majhi:2015tpa,Majhi:2017fua}.
To get the explicit expressions of this bracket and the charge in this theory, we shall follow the methods prescribed in \cite{Carlip:1999cy}.
We take a set of diffeomorphism generators defined by the relation $\xi^a=T\chi^a+R\rho^a $, where $\chi^{a}$ is the timelike Killing vectors and $\rho^{a}$ is the orthogonal to the Killing vector. The component $R$ of the diffeomorphism generator $\xi^{a}$ along the orthogonal vector $\rho^{a}$ is related [as argued in \cite{Majhi:2011ws}, Eq. (12)] by $R=\frac{\chi^2}{\kappa\rho^2}DT$, where $D \equiv\chi^{a}\nabla_{a}$. Now, the lie bracket of the diffeomorphism operators is given by (Eq. (26) of \cite{Majhi:2011ws}),
\begin{equation}
\{\xi_1,\xi_2\}^a=(T_1DT_2-T_2DT_1)\chi^a -\frac{1}{\kappa}D(T_1DT_2-T_2DT_1)\rho^a \label{XIXI}
\end{equation}

For the mentioned set of diffeomorphism generators, one can calculate (for details, see Appendix \ref{APPEN2})
\begin{align}
Q[\xi]=\frac{1}{16\pi}\int \sqrt{h}d^2x\phi(2\kappa T-\frac{1}{\kappa}D^2T) \label{QAPP}
\end{align}
and
\begin{eqnarray}
&&[Q_1,Q_2]:=\frac{1}{16\pi}\int\sqrt{h}d^2x\phi[2\kappa(T_1DT_2-T_2DT_1)
\nonumber
\\
&&-\frac{1}{\kappa}(T_1D^3T_2-T_2D^3T_1)] \label{BRAC12}
\end{eqnarray}
Using the relation \eqref{XIXI}, one obtains
\begin{eqnarray}
&&Q[\{\xi_1,\xi_2\}]=\frac{1}{16\pi}\int\sqrt{h}d^2x[2\kappa(T_1DT_2-T_2DT1)
\nonumber
\\
&&-\frac{1}{\kappa}D^2(T_1DT_2-T_2DT_1)]\phi \label{AB12} 
\end{eqnarray}
Therefore, the central term, defined by the relation $K[\xi_1,\xi_2]=[Q_1,Q_2]-Q[\{\xi_1,\xi_2\}]$, is
\begin{align}
K[\xi_1,\xi_2]=\frac{1}{16\pi}\int \sqrt{h}d^2x \phi\frac{1}{\kappa}[(DT_1)(D^2T_2)-(DT_2)(D^2T_1)] \label{KULT}
\end{align}
If one takes the Fourier modes of $T$ as $T={\sum_{m}}A_mT_m$ and 
takes the usual ansatz for $T_m$ as given in \cite{Majhi:2011ws},
\begin{align}
T_m=\frac{1}{\kappa}Exp[i m\{\kappa t+g(x)+p.x_{\bot}\}], \label{FOUT}
\end{align}
then the Fourier modes of the charges are
\begin{align}
Q[\xi_m]=\frac{\int\sqrt{h}d^2x\phi}{8\pi}\delta_{m,0}=\frac{A\phi}{8\pi}\delta_{m,0} \label{FOUQ}
\end{align}
where we define $\int \sqrt{h}d^2x=A$ as the area of the null surface. We want to comment on the fact that $\phi$ is the function of all coordinates in general. As the calculations in this section are near the horizon, we expand $\phi$ about the horizon to get $\phi(t,r,X^A)=\phi(r_H)+(r-r_H)\phi^{'}(t,r_H,X^A)+..$. Near the horizon only the first term contributes which is independent of all the coordinates to be consistent with the zeroth law of thermodynamics \cite{Faraoni:2010yi}. Therefore, the $\phi$ in the Eq. \eqref{FOUK} is actually the first term of the expansion of $\phi(t,r,X^A)$. We shall use this convention throughout this section.
Similarly, one obtains
\begin{align}
Q[\{\xi_m,\xi_n\}]=i(m-n)Q[\xi_{m+n}].
\end{align}
and
\begin{align}
K[\xi_m,\xi_n]=-\frac{im^3\phi A}{8\pi}\delta_{m+n=0} \label{FOUK}
\end{align}

From \eqref{AB12} and \eqref{KULT}, collecting all the terms, we obtain
\begin{align}
i[Q_m,Q_n]=(m-n)Q[\xi_{m+n}]+ m^3\frac{A\phi}{8\pi}\delta_{m+n,0} \label{FOUBRA}
\end{align}

Comparing the obtained relation with the stranded Virasoro  algebra of charges given by
\begin{align}
i[Q_m,Q_n]=(m-n)Q_{m+n}+\frac{C}{12}m^3\delta_{m+n,0}, \label{VIR}
\end{align}
one finds
\begin{align}
\frac{C}{12}=\frac{A\phi}{8\pi} \label{COMPARE}
\end{align}
One already knows that the Cardy formula for entropy is given by the relation \cite{Carlip:1998qw,Carlip:1999cy}
\begin{align}
S=2\pi\sqrt{\frac{C\Delta}{6}} \label{CARDY}
\end{align}
where $\Delta$ is defined as $\Delta=Q_0-\frac{C}{24}$, $Q_0=\frac{A\phi}{8\pi}$ being the zeroth mode of charge. Therefore, the expression of entropy determined by the Cardy formula is given by
\begin{align}
S=\frac{A\phi}{4} \label{ENTROPY}
\end{align}
We see that the expression of entropy $S=\frac{\phi A}{4}$ agrees to the Kang's prescription \cite{Kang:1996rj}.
\\
 For the Einstein frame, the calculation will be exactly similar to the usual GR case \cite{Majhi:2011ws} as the Noether current is exactly the same in both the cases. So, in that case, $\tilde{S}=\frac{\tilde{A}}{4}$, where $\tilde{A}=\int\sqrt{\tilde{h}}d^2x=\phi A$. Therefore, the entropy in both the frame are the same.

 So, using the Virasoro algebra, we find the entropy of the two frames to be equivalent. We proved this without any assumption and without including any extra prescription by hand. The equivalence of the entropy in the two frames is used later in this paper to obtain the relation of the other thermodynamic quantities (the energy and the temperature) in the two frames.

\textit{\textbf{Summary:} We obtain the expression of the entropy from the first principal, it is proved that the entropy in the two frames are equivalent. For the theory in the Jordan frame, the entropy is proportional to the area of the surface horizon as well as the scalar field $\phi$.}

\section{Comparison at the thermodynamic level}
In the following part we shall find the correspondence of the thermodynamic variables between the two frames.  Although in \cite{Koga:1998un}, it has been proved that all the thermodynamic quantities are equivalent. But, it is based on a few assumptions along with the another as the two vectors $\xi^a$ and $\tilde{\xi}^a$ to be the same without any justification. In the previous section we obtained the expressions of the entropy in the two frames and we proved they are equivalent in the two frames without any assumption. Now we try to develop the relation of energy and temperature in the two frames. Our approach is completely different from all the others in the literature and it is not based on any assumption. Instead, our analysis will prove that $\tilde{\xi}^a=\xi^a$, justifying the earlier assumption of \cite{Koga:1998un} and our results also matches to it.

\subsection{Action as the free energy of spacetime} 
One of the most interesting fact of the Einstein-Hilbert action is that it can be treated as the free energy of spacetime for stationary background (For details see \cite{Padmanabhan:2004fq}). The question that naturally arises is whether the same interpretation of the action is applicable for this theory or not. Firstly, we try for this theory in Einstein frame, as the action quite resembles to the Einstein-Hilbert one and then we do for the theory in Jordan frame.

$\bold{Einstein \ frame:}$ The gravitational action in this frame is given by \eqref{SE}. Now the equation of motion of the metric tensor in this frame, when the matter field is included, is given in \eqref{EOM1}, and only the  zero on the right-hand side is replaced by $\frac{\tilde{T}_{ab}}{2}$, where, $\tilde{T}_{ab}$ is the energy-momentum tensor of the matter field in the Einstein frame. If one rises one index of the equation of motion and then fixes the value of both indexes as zero, one obtains $\tilde{G}_0^0=\tilde{R}_0^0-\frac{1}{2}\tilde{R}=\frac{1}{2}\tilde{\nabla}^0\tilde{\phi}\tilde{\nabla}_0\tilde{\phi}-\frac{1}{4}\tilde{\nabla}^i\tilde{\phi}\tilde{\nabla}_i\tilde{\phi}-\frac{U}{2}+\frac{\tilde{T}^0_0}{2}$. From this relation if one replaces $\tilde{R}$ in the Lagrangian given in \eqref{SE}, one obtains 
\begin{align}
&&\tilde{L}=-\tilde{T}_0^0+\frac{2\tilde{R}_0^0}{16\pi}-\tilde{\nabla}_0\tilde{\phi}\tilde{\nabla}^0\tilde{\phi} \label{FRE1}
\end{align}
The last term in the above equation does not contribute as we have assumed the spacetime is stationary and therefore, the field $\phi$ is independent of the time.
It should be mentioned here that the components of the Energy-Momentum tensor of the Einstein fame is related to the Jordan frame with the relation $\tilde{T}_a^b=\frac{T_a^b}{\phi^2}$. The stationary spacetime has a timelike Killing vector $\tilde{\chi}^a=(1,0,0,0)$. Therefore, a straightforward calculation gives 
\begin{align}
\tilde{R}^a_j\tilde{\chi}^j=\tilde{\nabla}_b\tilde{\nabla}^a\tilde{\chi}^b=\frac{1}{\sqrt{-\tilde{g}}}\partial_b(\sqrt{-\tilde{g}}\tilde{\nabla}^a\tilde{\chi}^b), \label{RXI}
\end{align}
where, the last identity comes from the fact that $\tilde{\nabla}^a\tilde{\chi}^b$ is an antisymmetric tensor (following the relation of the Killing vector). Since $\tilde{\chi}^a$ has only the time component, one obtains
\begin{align}
R_0^0=\frac{1}{\sqrt{-\tilde{g}}}\partial_b(\sqrt{-\tilde{g}}\tilde{g}^{i0}\tilde{\Gamma}^b_{i0}) \label{R00}
\end{align}
Note, that the index $b$ cannot be time as the spacetime is stationary. Moreover, one should integrate the Lagrangian in the four-dimensional spacetime manifold, and for that, one has to take finite range of time (0,$\tilde{\beta}$) to get a finite result as the spacetime is the stationary one. Therefore, one can express the action as 
\begin{align}
\tilde{\mathcal{A}}=\tilde{\beta}\int\tilde{N}\sqrt{\tilde{h}}\tilde{\rho}d^3\tilde{x}+\frac{\tilde{\beta}}{8\pi}\int d^2\tilde{x}\sqrt{\tilde{\sigma}}\tilde{N}\tilde{n}_{\alpha}(\tilde{g}^{i0}\tilde{\Gamma}^{\alpha}_{i0}) \label{FREE1}
\end{align}
The last term is obtained after converting the space volume integral of $\tilde{R}_0^0$ to the surface integral. Here, $\tilde{h}$ and $\tilde{\sigma}$ are the determinant of induced 3- and 2-metric respectively, and  $\tilde{N}=\sqrt{-\tilde{g}_{00}}$ is the lapse function. We have also used the relation $\tilde{T}^0_0=-\tilde{\rho}$. Thus, identifying the integral part of the first term of  \eqref{FREE1} as the energy and the whole second term as the (negative of) entropy one gets $\tilde{\mathcal{A}}\equiv \tilde{\beta}\tilde{E}-\tilde{S}$. So, one can conclude that the action can be interpreted as the free energy of the spacetime. Earlier the same interpretation was given for the GR as mentioned earlier and one find that the same interpretation is applicable for this scalar-tensor theory in the Einstein frame as well.\\
Let us now find out whether one can give the same inference of the action of this theory in Jordan frame.

{\bf Jordan frame:} The gravitational action in this frame is given by \eqref{SJ}. In this case as well, one should use the equation of motion of the metric tensor \eqref{EOM3}. One has to rise one index and then one has to fix both indexes as zero (as done in the Einstein frame) to obtain
\begin{align}
&&L=\frac{1}{16\pi}(2\phi R^0_0+16\pi\rho+\frac{2\omega}{\phi} \nabla^0\phi\nabla_0\phi-2\nabla^0\nabla_0\phi+
\nonumber
\\
&&2\square\phi) \ \ \ \ \ \ \ \ \ \ \ \ \ \ \ \ \ \ \ \ \ \ \ \ \ \ \ \ \ \ \ \ \ \ \ \ \ \ \ \ \ \ \ \ \ \ \ \ \ \ \ \ \ \label{FREE2}
\end{align}
 Now, in general, for any vector one can write
\begin{align}
 R_j^b(\phi\chi^j)=\nabla_a\nabla^b(\phi\chi^a)-\nabla^b\nabla_a(\phi\chi^a) \label{RICCI}
\end{align}
For this theory in stationary background, let there be the timelike Killing vector $\chi^a=(1,0,0,0)$.
Now, the second term $\nabla^b\nabla_a(\phi\chi^a)=\nabla^b(\chi^a\nabla_a\phi)-\nabla_b(\phi\nabla_a\chi^a)$ vanishes using the property of the Killing vector $\chi^a$. So, 
\begin{align}
\phi R_j^b\chi^j=\nabla_a\nabla^b(\phi\chi^a) \ \ \ \ \ \
\nonumber
\\
=\nabla_a[\phi(\nabla^b\chi^a)]+(\nabla_a\chi^a)(\nabla^b\phi)+\chi^a[\nabla_a\nabla^b\phi] \label{RICCI1}
\end{align}
The first term in the above equation is an antisymmetric tensor which can be written in terms of a total derivative form. The second term vanishes and therefore one obtains
\begin{align}
\phi R_j^b\chi^j=\frac{1}{\sqrt{-g}}\partial_a[\phi\sqrt{-g}(\nabla^b\chi^a)]+\chi^a\nabla_a\nabla^b\phi \label{RICCI2}
\end{align}
As, the Killing vector has only the time component, one obtains
\begin{align}
\phi R^0_0-\nabla_0\nabla^0\phi=\frac{1}{\sqrt{-g}}\partial_a[\phi\sqrt{-g}g^{0i}\Gamma^a_{i0}] \label{RICCI3}
\end{align}
So, ultimately one can obtain
\begin{eqnarray}
&&\mathcal{A}=\frac{1}{16\pi}\int \sqrt{-g}d^4x \big{[}16\pi\rho-\frac{2}{\sqrt{-g}}\partial_a[\phi\sqrt{-g}g^{0i}\Gamma^a_{i0}]
\nonumber
\\
&&+2\square\phi\big{]}
\nonumber
\\
&&=\frac{1}{16\pi}\int \sqrt{-g}d^4x \big{[}16\pi\rho-\frac{2}{\sqrt{-g}}\partial_a[\sqrt{-g}(\phi g^{0i}\Gamma^a_{i0}
\nonumber
\\
&&+g^{aj}\partial_j\phi)\big{]}
\end{eqnarray}
As done earlier, here also the action can be treated as the free energy of the spacetime
\begin{eqnarray}
&&\mathcal{A}=\beta\int N\sqrt{h}d^3x\rho
\nonumber
\\
&&+\frac{\beta}{8\pi}\int d^2x\sqrt{\sigma}Nn_{\alpha}(\phi g^{0i}\Gamma^{\alpha}_{0i}+g^{\alpha j}\partial_j\phi)
\nonumber
\\
&&=\beta E-S. \label{FREE3}
\end{eqnarray}
In this frame, the second integration gives (the negative of) the entropy.\\
  A few comments should be made in this regards. Earlier we have shown that the entropy in the two frame are the same. Again, one can show 
\begin{align}
&&\sqrt{\tilde{\sigma}}\tilde{N}\tilde{n}_{\alpha}(\tilde{g}^{i0}\tilde{\Gamma}^{\alpha}_{i0})=\sqrt{\sigma}Nn_{\alpha}(\phi g^{i0}\Gamma^{\alpha}_{i0})-\frac{1}{2}\sqrt{\sigma}Nn_{\alpha}g^{\alpha j}\partial_j\phi \label{VB}
\end{align}
Now, $\phi=\Omega^2$ is not the function of time. Also, the last term on the right-hand side vanishes at the boundary of a stationary spacetime. So, one gets $\sqrt{\tilde{\sigma}}\tilde{N}\tilde{n}_{\alpha}(\tilde{g}^{i0}\tilde{\Gamma}^{\alpha}_{i0})=\sqrt{\sigma}Nn_{\alpha}(\phi g^{i0}\Gamma^{\alpha}_{i0})$. Therefore, for a stationary metric one can show (from the expressions of the entropy in the two frames) $\tilde{\beta}=\beta$ as the entropy is the same in the two frames.
Again, one can also prove $\tilde{N}\sqrt{\tilde{h}}\tilde{\rho}=N\sqrt{h}\rho$, as $\tilde{N}=\sqrt{-\tilde{g}^{00}}=\sqrt{\phi}N$, $\sqrt{\tilde{h}}=\phi^{\frac{3}{2}}\sqrt{h}$ and $\tilde{\rho}=\tilde{T}^0_0=\frac{T^0_0}{\phi^2}=\frac{\rho}{\phi^2}$. 
\vskip 1mm
\textit{\textbf{Summary:} Like the analysis done in the action level, it has been shown here that the total action in both the frames can be separated out in the two parts. The total derivative one, which we identify as the surface term in the thermodynamic level and accounts to the entropy in both the frames. Also, there is the term containing the component of the energy-momentum tensor which gives the energy in both the frames. Later, we have also shown that for the stationary metric, the temperature and the energy is invariant in the two frames.}
\subsection{Holographic relation at the thermodynamic level}
The holographic relation was obtained earlier in the action level. We found that in the Einstein frame the holographic relation holds but in the Jordan frame a total derivative term blemishes the relation. Here we try to obtain the relation in the thermodynamic level and find that a total derivative term in both the frames accounts for the entropy, and the remaining term contributes as the energy. Let us call the earlier one the surface term ($L_{sur}$) and the later one, the bulk term ($L_{bulk}$)\footnote{Note, in general, that the bulk or surface terms in the two cases (the analysis in the classical level and here in the thermodynamic level) are not the same.}. Our aim is now to see if we can have a similar relation like in the earlier analysis.

To proceed, here one trick will be used followed from \cite{Kolekar:2010dm}. Firstly one has to calculate the variation of total Lagrangian (equivalent to $L_2$($q$, $\partial q$, $\partial^2q$) of \eqref{REL}) and $L_(sur)$. Then the difference of this two variation gives the same for the bulk part $L_{bulk}$. Note that the total Lagrangians [ \eqref{SJ} and \eqref{SE}] in the two frames are the function of the metric tensor, its first derivative and the second derivative and the scalar field and its first-order derivative. Therefore, following the earlier logic of the section \ref{HOL}, we shall consider the variation with respect to the metric tensor and its derivatives (first and second order) only.
\vskip 1mm
$\bold{Einstein \ frame:}$ 
We have identified the surface term in the thermodynamic level from \eqref{FRE1} as $\tilde{L}_{sur}=\frac{2\tilde{R}_0^0}{16\pi}$. Therefore, writing the bulk part in the thermodynamic level as  $\tilde{L}_{bulk}=\tilde{L}-\tilde{L}_{sur}$ and following the calculations of Appendix A of \cite{Kolekar:2010dm} one can show, 
\begin{eqnarray}
&& \tilde{L}_{sur}=-\partial_i\Big(\tilde{g}_{ab}\frac{\delta_E \tilde{L}_{bulk}}{\delta_E (\partial_i \tilde{g}_{ab})}+\partial_j\tilde{g}_{ab}\frac{\partial \tilde{L}_{bulk}}{\partial(\partial_i\partial_j\tilde{g}_{ab})}\Big)~, \label{7}
\end{eqnarray} 
where $\frac{\delta_E \tilde{L}_{bulk}}{\delta_E (\partial_i \tilde{g}_{ab})}$ implies the Euler derivative of the bulk Lagrangian with respect to $\partial_ig_{ab}$.  Thus one can conclude that in the thermodynamic level as well one can obtain the holographic relation for the theory in the Einstein frame.

$\bold{Jordan \ frame:}$
Here the total Lagrangian is given in \eqref{SJ} and the surface part, from \eqref{FREE2}, is identified as, $L_{sur}=\frac{1}{16\pi}(2\phi R_0^0-2\nabla^0\nabla_0\phi+2\square\phi)$. So, the bulk part in the thermodynamic level is $L_{bulk}=L-L_{sur}$.
 We obtain (for detailed calculation, see the Appendix \ref{appp}) 
\begin{eqnarray}
&& \partial_i \Big(g_{ab}\frac{\delta_E L_{bulk}}{\delta_E(\partial_i g_{ab})}+\partial_j g_{ab}\frac{\partial L_{bulk}}{\partial(\partial_j \partial_i g_{ab} )} \Big)=-L_{sur}
\nonumber
\\
&&+\frac{3}{16}\sqrt{-g}\square\phi \label{8}
\end{eqnarray}
 So, in the thermodynamic level one does not get the holographic relation in Jordan frame as was the case in the action level earlier. Moreover, one can check that it is the same extra term that spoils the holographic relation. 
 \vskip 1mm
  \textit{\textbf{Summary:} Like the analysis done in the action level, here we have shown that the surface and the bulk part are connected by the holographic relation in the Einstein frame. But, one cannot draw the same conclusion for the theory in the Jordan frame. The same extra term appears that blemishes the theory to be holographic in the latter frame as was the case in the analysis in the action level.}
\subsection{Relation between different thermodynamic entities from the GHY term}
So far, all the calculations are done for the gravitational actions without the GHY term. It has been argued that the thermodynamic entities are invariant for the stationary background.  In this section, we want to justify our arguments again from the GHY boundary term. 

The GHY surface term in Einstein frame, mentioned in \eqref{SEGHY}, provides the antisymmetric Noether potential (see the Appendix of \cite{Majhi:2012tf} )
\begin{align}
\tilde{J}^{ab}_S=\tilde{K}(\tilde{N}^a\tilde{\xi}^b-\tilde{N}^b\tilde{\xi}^a)~. \label{JABE}
\end{align}
Similarly, the GHY term in the conformally connected Jordan's frame, mentioned in \eqref{SJGHY} yields the Noether potential
\begin{align}
J^{ab}_S=\Omega^2 K(N^a\xi^b - N^b\xi^a)~. \label{JABJ}
\end{align}
To develop the relation between the thermodynamic quantities, we assume that the diffeomorphism vectors $\tilde{\xi}^a$ and $\xi^a$ of the two frames are proportional to each other with the proportionality constant being the arbitrary power of the conformal factor. Let,
\begin{align}
\tilde{\xi}^a=\Omega^{\alpha}\xi^a~, \label{XI}
\end{align}
where, $\Omega=\sqrt{\phi}$ in this theory. Unlike \cite{Koga:1998un} we do not consider that the two vectors are equal. It should be mentioned here that in \cite{Koga:1998un}, they have made the mentioned assumption and thereby they have shown that the thermodynamic quantities are the same in the two frames.
Here, in this discussion we want to fix the value of $\alpha$ and from that we shall show that we can get the relationship between the thermodynamic quantities of the two frames. Coincidentally, we shall show that our result matches theirs.

Using the relations \eqref{K} and \eqref{XI}, one can easily obtain
\begin{align}
\tilde{J}^{ab}_S=\Omega^{\alpha -4}J^{ab}_S-3(N^i\partial_i\Omega)\Omega^{\alpha-3}(N^a\xi^b-N^b\xi^a)~. \label{JAB}
\end{align}
Now, following the usual process, if one defines the conserved Noether charges as (in Einstein frame and in Jordan frame) $\tilde{Q}=\frac{1}{2}\int d\tilde{\Sigma}_{ab}\tilde{J}^{ab}_S$, $Q=\frac{1}{2}\int d\Sigma_{ab}J^{ab}_S$ then for a stationary black hole ($\xi^a=\chi^a$ and $\tilde{\xi}^a=\tilde{\chi}^a$, where $\chi^a$ and $\tilde{\chi}^a$ are the time-like Killing vectors) one obtains
\begin{align}
\tilde{Q}=\Omega^{\alpha}Q \label{Q}
\end{align}
on the horizon. To achieve this one needs to use the relations $d\tilde{\Sigma}_{ab}=(\tilde{l}_a\tilde{\chi}_b-\tilde{l}_b\tilde{\chi}_a)d^2x_{\bot} \sqrt{\tilde{\sigma}}$, $d\Sigma_{ab}=(l_a\chi_b-l_b\chi_a)d^2x_{\bot} \sqrt{\sigma}$, $\tilde{l}_a=\Omega^{-\alpha}l_a$ and $\sqrt{\tilde{\sigma}}=\Omega^2\sqrt{\sigma}$. Here, $l_a$ and $\tilde{l}_a$ are the auxiliary null vectors in the two frames satisfying $\tilde{\chi}^a\tilde{l}_a=-1=\chi^al_a$. 

From the Noether potentials \eqref{JABE} and \eqref{JABJ} if one calculates the entropy on a null surface, one defines entropy in, say Jordan frame as $S=\frac{2\pi}{\kappa}Q$. One uses the similar definition in Einstein frame as well. The surface gravity $\kappa$ is defined by the relation $\chi^a\nabla_a\chi^b=\kappa\chi^b$. For the transformation mentioned in \eqref{XI} one can find $\tilde{\kappa}=\Omega^{\alpha}\kappa$ with $\tilde{\kappa}$ defined as $\tilde{\chi}^a\tilde{\nabla}_a\tilde{\chi}^b=\tilde{\kappa}\tilde{\chi}^b$. Therefore, from \eqref{Q} one finds that the entropy is the same in the two frame. This implies that for any value of $\alpha$, the entropy is equivalent in the two frames. So, it is not surprising that \cite{Koga:1998un} got the same conclusion  as the entropy is invariant in the two frames with the specific choice of $\alpha=0$. But, for the other thermodynamic quantities one has to fix $\alpha$ to get the proper relation in the two frames. For instance,
\begin{align}
\tilde{T}=\Omega^\alpha T~,\label{TEMP}
\end{align}
where $T=\frac{\kappa}{2\pi}$ and so on. To fix $\alpha$ we take a particular example.
Consider a static, spherically symmetric(SSS) metric in the Jordan frame \footnote{We are not going into the arguments in some paper (for example \cite{Faraoni:2016ozb}) whether the SSS metric represents a black hole solution in the scalar-tensor theory. Here, the calculations are done on the null surface}
\begin{align}
ds^2=f_1(r)dt^2-\frac{dr^2}{f_2(r)}-r^2(d\theta^2+\sin^2\theta d\Phi^2) \label{MET1}
\end{align}
Now, the GHY surface term in Jordan frame, mentioned in \eqref{SJGHY} gives contribution on r=const  surface as 
\begin{align}
\mathcal{A}_{GHY}\mid_r=-\frac{1}{8\pi}\int_{r=r_H}\phi\sqrt{h^{(r)}}K^{(r)}d^3x
\nonumber
\\
=-\frac{\beta\sqrt{f_1^{'}(r_H)f_2^{'}(r_H)}}{4}(\phi r_H^2)=-\phi \pi r_H^2=-S \label{RJOR}
\end{align}
where, $T=\frac{\kappa}{2\pi}=\frac{\sqrt{f_1^{'}(r_H)f_2^{'}(r_H)}}{4\pi}=\beta^{-1}$ from the definition of surface gravity has been used. We have defined $\phi$ times quarter of the surface area as the entropy. Here,  $N_r=\frac{1}{\sqrt{\mid g^{rr}}\mid}=\frac{1}{\sqrt{f_2(r)}}$ is the normal defined in the r=const hypersurface .\\
Let us recall the fact mentioned in the section \ref{EOM} that as the equations of motion are equivalent in the two frames by the transformations \eqref{GAB} and \eqref{PHI}.  So if the mertic $g_{ab}$ is the solution in the Jordan frame, then $\phi g_{ab}$ is the solution in Einstein frame. Hence, the corresponding conformally connected metric tensor of \eqref{MET1} is the solution in the Einstein frame:
\begin{align}
d\tilde{s}^2=\phi\Big[f_1(r)dt^2-\frac{dr^2}{f_2(r)}-r^2(d\theta^2+\sin^2\theta d\Phi^2)\Big] \label{MET2}
\end{align}
The same calculation (calculation of the GHY surface term \eqref{SEGHY} at constant $r$ surface) in Einstein's frame leads to the result 
\begin{align}
&&\tilde{\mathcal{A}}_{GHY}\mid_r=-\frac{1}{8\pi}\int_{r=r_H}\sqrt{\tilde{h}^{(r)}}\tilde{K}^{(r)}d^3x
\nonumber
\\
&&=-\frac{\tilde{\beta}\sqrt{f_1^{'}(r_H)f_2^{'}(r_H)}}{4}(\phi r_H^2)
\nonumber
\\
&&=-\tilde{S}, \ \ \ \ \ \ \ \ \ \ \ \ \ \ \ \ \ \ \ \ \ \ \ \ \ \ \ \ \label{REIN}
\end{align}
where, $\tilde{N}_r=\sqrt{\frac{\phi}{f_2(r)}}$.
One can verify that the two surface actions $\mathcal{A}_{GHY}$ and $\tilde{\mathcal{A}}_{GHY}$ given in \eqref{SEGHY} and \eqref{SJGHY} are invariant near the null surface as they are connected with each other by the relation \eqref{K} and the last term of that equation does not contribute at the r=const surface. Therefore, we can say the result of \eqref{RJOR} and \eqref{REIN} are the same. From that, one can conclude that $\tilde{\beta}=\beta$. Hence, the temperature is invariant in both the frames. So, from \eqref{TEMP} one gets $\alpha=0$ which justifies the assumption of \cite{Koga:1998un}.


Next, let us see how the gravitational energy are related in the two frames. For that, let us calculate the GHY terms in all the frames. The motivation is from the fact that in the GR case the surface term is interpreted as the free energy of the spacetime($\mathcal{A}_{GHY}^{(GR)}=-S^{(GR)}+\beta^{(GR)}E^{(GR)}$) when it is calculated for all the surfaces (for details, see \cite{Padmanabhan:2002sha}). Note that when the GHY term is calculated for all the surfaces, generally there are four surfaces constant $r$, constant $\theta$, constant $\phi$ and constant $t$. But, for SSS spacetime, the contribution from the last two surfaces vanishes as the metric components are independent of $t$ and $\phi$.
The GHY surface term, when calculated collectively on all the surfaces, gives 
\begin{align}
&&\mathcal{A}_{GHY}=\mathcal{A}_{GHY}\mid_r+\mathcal{A}_{GHY}\mid_{\theta}
\nonumber
\\
&&=-S-\frac{1}{8\pi}\int_{\theta}\phi\sqrt{h^{(\theta)}}K^{(\theta)}d^3x
\nonumber
\\
&&=-S+\frac{\beta}{2}\int_0^{r_H}\phi\sqrt{\frac{f_1}{f_2}}dr=-S+\frac{E}{T}, \label{THETAJ}
\end{align}
where $E$ is defined as $E=\frac{1}{2}\int_0^{r_H}\phi\sqrt{\frac{f_1}{f_2}}dr$\\
and the same calculation in Einstein frame gives
\begin{align}
\tilde{\mathcal{A}}_{GHY}=\tilde{\mathcal{A}}_{GHY}\mid_r+\tilde{\mathcal{A}}_{GHY}\mid_{\theta}
\nonumber
\\
=-\tilde{S}-\frac{1}{8\pi}\int_{r=r_H}\sqrt{\tilde{h}^{(\theta)}}\tilde{K}^{(\theta)}d^3x
\nonumber
\\
=-\tilde{S}+\frac{\tilde{\beta}}{2}\int_0^{r_H}\phi\sqrt{\frac{f_1}{f_2}}dr=-\tilde{S}+\frac{\tilde{E}}{\tilde{T}}. \label{THETAE}
\end{align}
Like the GR cases, we found that the GHY term has the free energy structure in both the frames.
Now, one can show that two GHY actions are the same on the horizon as they are connected by the relation \eqref{K}, and the last term of that equation does not contribute there. So we can say the results  of\eqref{RJOR} and \eqref{REIN} are the same.
Since $\tilde{T}=T$ and $\tilde{S}=S$, one gets the equivalent expression of the energy in both frames as $\tilde{E}=E=\frac{1}{2}\int_0^{r_H}\phi\sqrt{\frac{f_1}{f_2}}dr$.

 Now our analysis suggests that all the thermodynamic variables are invariant in both frames. Reference \cite{Jacobson:1993pf} argues that the expression of the temperature should be equivalent under the conformal transformation. Our arguments give identical results, although obtained in a different fashion. The literature \cite{Koga:1998un} demands that all the thermodynamic variables are invariant in both the Jordan and the Einstein frame when the spacetime is asymptotically flat. To prove these, the authors made the key assumption that the vectors $\tilde{\xi}^a$ and $\xi^a$ are the same in both the frames. There is no justification for how they took it for granted. Our analysis gives the identical results, although it is much more robust.
 
 It should be mentioned here that although, the energy in the two frames should be invariant, but, unfortunately, we have not been able to give any covariant form of the energy. So, we cannot compare it with existing expression of energy in the literature. But, we can predict whatever may be the expression, it should be conformally invariant. Interestingly, one candidate fulfils this condition on the horizon which is the Brown-York energy \cite{Brown:1992br}. On the other hand, the recent works suggests that the Misner-Sharp energy \cite{Faraoni:2014lsa} or the Hawking-Heyward quasilocal energy (\cite{Prain:2015tda}, \cite{Faraoni:2015sja}, \cite{Hammad:2016yjq})  and the others are not conformally invariant and, hence, they cannot be the candidates. Although, \cite{Bose:1998yp} suggests that the Brown York energy is conformally invariant all over the spacetime but that is not correct. The wrong statement of the mentioned paper is due to the fact that the Eq. (2.25) of \cite{Bose:1998yp} (the relation between the traces of the two extrinsic curvature tensors in the two frames) is incorrect. For (1+3) dimensions, we derived the correct one in \eqref{K}. This correct one tells us that the Brown-York energy is the same only on the horizon.

Thus, in our analysis, we found that the thermodynamic quantities are invariant in the two frames from the whole action as well as from the GHY surface term. So, from the thermodynamic point of view, the two frames are equivalent for the static spherical symmetric spacetime.

\vskip 1mm
\textit{\textbf{Summary:} The GHY term in both the frames can be interpreted as the free energy of the spacetime. Moreover, only from the surface part one can conclude that the thermodynamic quantities (entropy, energy and temperature) are invariant in both the frames.}

\section{Conclusion}
In this work, we have deeply studied the well-known scalar-tensor theory from action as well as thermodynamic framework. Also, some consideration, which is often taken for granted without even proper declaration, is mentioned here. Moreover, the work also assists for a comparative study of this theory with the usual GR one. It has earlier been mentioned that the argument whether the two frames, on which the theory is described, is equivalent is not still been resolved and people are studying the two frames in numerous ways to distinguish the equivalence (or inequivalence) of the two frames. This work basically have addressed those issues, though studied from the different perspective.

 Here, we have started analyzing the theory in the two frames form the action level. What we found is the usually mentioned mathematical equivalence in the literature is an incomplete statement. One just unseeingly neglects a total derivative term while projecting the theory from one frame to the other. It has been shown that the concern does not arise while one incorporates the GHY surface term in this theory. Which means, the usual description of the mathematical equivalence of the two frames breaks while one does not include the GHY surface term. After that, we have separated the gravitational action in each frame as a bulk part and a total derivative surface part. Later the separation is justified by obtaining the equation of motion from the bulk part of the total action. Thereafter, we endeavored to obtain the connection of the two part of the gravitational action by the holographic relation. Unlike the GR case, it has been shown here that the holographic property is not valid in the original(Jordan) frame. But, it continues to be valid in the Einstein one and, thereby, obtaining the inequivalence  of the two frames in the action (classical) level even without the presence of the external matter field.

Afterwards, the two frames are compared at the thermodynamic level. The relation of the thermodynamic entities in the two frames were not well known. It has also been the subject of debate which form of the energy in the literature should be used to describe the thermodynamics of the system in this theory. Although, some earlier attempts has been made to solve the enigma, those are based on some assumptions. In this paper, the entropy in the two frames has been obtained from the first principle using the Virasoro algebra technique, the first attempt to obtain entropy in this theory in that way. We have been able to prove that the entropy in the two frames are indeed equivalent. Later, we have interpreted the gravitational action as the free energy of the spacetime in the two frames and it has also been shown that the identified expression of the energy and the temperature in the two frames is equivalent. The result coincides with some previous works, though obtained in a more robust way. Also, the attempt has been made to get the ``holographic'' relation at the thermodynamic level as we have separated out the total action in the thermodynamic level as a bulk part which was identified in terms of energy and a total derivative surface term which accounts to the entropy. The result obtained here is the same as earlier; i.e., the holographic relation is maintained in the Einstein frame, while it is defiled in the original one. We have also noticed that the same term appears in this case as well to spoil the holographic property in the Jordan frame. Finally, all the earlier obtained relation of the thermodynamic entities has been verified from the GHY surface terms. Moreover, it has been shown that the GHY surface term itself can be interpreted as the free energy of the spacetime.

 The analysis suggests that the energy which should be used in this theory to describe the thermodynamics of the system, should be conformally invariant. The Brown-York formalism of the quasilocal energy in the literature is conformally invariant near the horizon. On the other hand, the Misner-Sharp energy or the Hawking-Heyward quasilocal energy are not conformally invariant. As the identified expression of energy in our analysis has not been obtained in a covariant form, we cannot count on any standard form of energy in the literature to be absolute one for the thermodynamic description. We only have found out that it should be conformally invariant. Therefore, it is more probable for the Brown-York one to describe the thermodynamics of the system in this theory rather than the Misner-Sharp or the Hawking-Heyward quasilocal one.
 
 Finally we want to make a comment that the method was used to identify the different thermodynamic quantities which lead to the fact that energy has to be conformally invariant. This is what we find within this method, but that there is no consensus on which method provides the ``right'' answer at this stage, so other approaches may provide different answers. The value of our analysis lies in the facts that it raises and in proposing what seems like an important approach, rather than in giving definitive answers. All the works in this regards (including this one) suggest that the investigation on these issues needs to be continued. So we hope we shall be able to give more insight in near future.

\vskip 4mm
{\section*{Acknowledgments}}
\noindent
We thank Valerio Faraoni for making various useful comments, suggestions and bringing important references to our attention during execution of the project.   
The research of one of the authors (BRM) is supported by a START-UP RESEARCH GRANT (No. SG/PHY/P/BRM/01) from Indian Institute of Technology
Guwahati, India.
\vskip 4mm
\appendix
\section{Derivation of eqs. \eqref{JAJ} and \eqref{JE1}} \label{APPEN1}
\subsection{Equation \eqref{JAJ}} \label{APPEN11}
The last two terms of Eq. \eqref{6} can be expressed as follows:
\begin{eqnarray}
&&2(\nabla_b\phi)P^{iabd}\pounds_{\xi} g_{id}
\nonumber
\\
&&=(\nabla^d\phi)[\nabla^a\xi_d+\nabla_d\xi^a]-2(\nabla^a\phi)(\nabla_i\xi^i) \label{A1}
\end{eqnarray}
and
\begin{eqnarray}
&& \pounds_{\xi}v^a=2P^{ibad}\nabla_b\pounds_{\xi}g_{id}=2P^{iabd}\nabla_b\pounds_{\xi}g_{id}
\nonumber
\\
&&=\nabla_b\nabla^a\xi^b+\nabla_b\nabla^b\xi^a-2\nabla^a\nabla_b\xi^b  \label{A2}
\end{eqnarray}
Therefore,
\begin{eqnarray}
&&2(\nabla_b\phi)P^{iabd}\pounds_{\xi} g_{id}-\phi\pounds_{\xi}v^a
\nonumber
\\
&&=(\nabla_b\phi)(\nabla^a\xi^b)+\phi\nabla_b\nabla^a\xi^b-\phi\square\xi^a 
\nonumber
\\
&&+(\nabla_b\phi)(\nabla^b\xi^a)-2(\nabla^a\phi)(\nabla_b\xi^b)-2\phi g^{ac}R_{kc}\xi^{k},
\end{eqnarray}
where one has to use the relation $\nabla_b\nabla_d\xi_i-\nabla_d\nabla_b\xi_i=R_{ijbd}\xi^j$. The first three terms of the above equation give $
(\nabla_b\phi)(\nabla^a\xi^b)+\phi\nabla_b\nabla^a\xi^b-\phi\square\xi^a=\nabla_b[\phi(\nabla^a\xi^b-\nabla^b\xi^a)]+(\nabla_b\phi)(\nabla^b\xi^a)
$ with this and using $2(\nabla_b\phi)(\nabla^b\xi^a)-2(\nabla^a\phi)(\nabla_b\xi^b)=2\nabla_b[\xi^a(\nabla^b\phi)-\xi^b(\nabla^a\phi)]+2\xi^b\nabla_b\nabla^a\phi-2\xi^a\square\phi$, one obtains
\begin{eqnarray}
&&2(\nabla_b\phi)P^{iabd}\pounds_{\xi} g_{id}-\phi\pounds_{\xi}v^a
\nonumber
\\
&&=\nabla_b[\phi(\nabla^a\xi^b-\nabla^b\xi^a)+2\xi^a(\nabla^b\phi)-2\xi^b(\nabla^a\phi)]
\nonumber
\\
&&+2\xi^b\nabla_b\nabla^a\phi-2\xi^a\square\phi-2\phi g^{ac}R_{kc}\xi^k
\end{eqnarray}
Using this in \eqref{6}, one can write Noether current as
\begin{eqnarray}
&&J^a=\frac{1}{16\pi}\Big[\nabla_b[\phi(\nabla^a\xi^b-\nabla^b\xi^a)+2\xi^a(\nabla^b\phi)-2\xi^b(\nabla^a\phi)]
\nonumber
\\
&&(\phi R-\frac{\omega (\phi)}{\phi}g^{ab}\nabla_a\phi \nabla_b\phi -V(\phi))\xi^a+2\frac{\omega}{\phi}(\nabla^a\phi)\xi^b (\nabla_b\phi)
\nonumber
\\
&&+2\xi^b\nabla_b\nabla^a\phi-2\xi^a\square\phi-2\phi g^{ac}R_{kc}\xi^k
\Big].
\end{eqnarray}
Now, using the equation of motion of the metric tensor $g^{bc}$ in this frame as given in \eqref{EOM3}, and then contracting it with $g^{ab}\xi^c$ one can obtain $(\phi R-\frac{\omega (\phi)}{\phi}g^{ab}\nabla_a\phi \nabla_b\phi -V(\phi))\xi^a+2\frac{\omega}{\phi}(\nabla^a\phi)\xi^b (\nabla_b\phi)+2\xi^b\nabla_b\nabla^a\phi-2\xi^a\square\phi-2\phi g^{ac}R_{kc}\xi^k=0$. Thus, the on shell Noether current is given by \eqref{JAJ}.
\subsection{Derivation of Eq. \eqref{JE1}} \label{APPEN12}
In Einstein frame, the Noether current is given by \eqref{JE}. Now, similar to the Jordan frame \eqref{A2} one can obtain
\begin{eqnarray}
&&\pounds_{\xi}\tilde{v}^a=\tilde{\nabla}_b\tilde{\nabla}^a\tilde{\xi}^b+\tilde{\nabla}_b\tilde{\nabla}^b\tilde{\xi}^a-2\tilde{\nabla}^a\tilde{\nabla}_b\tilde{\xi}^b
\nonumber
\\
&&=\tilde{\nabla}_b\tilde{\nabla}^b\tilde{\xi}^a-\tilde{\nabla}_b\tilde{\nabla}^a\tilde{\xi}^b+2\tilde{g}^{ac}\tilde{R}_{kc}\tilde{\xi}^k \label{A7}
\end{eqnarray}
To obtain the last step one has to use the same relation $\nabla_b\nabla_d\xi_i-\nabla_d\nabla_b\xi_i=R_{ijbd}\xi^j$ in the tilde frame. So, using \eqref{A7} one can obtain 
\begin{eqnarray}
&&\tilde{J}^a=\Big[(\frac{\tilde{R}}{16\pi}-\frac{1}{2}\tilde{g}^{ij}\tilde{\nabla}_i\tilde{\phi}\tilde{\nabla}_j\tilde{\phi}-U(\tilde{\phi})\Big]\tilde{\xi}^a+(\tilde{\nabla}^a\tilde{\phi})\tilde{\xi}^b(\tilde{\nabla}_b\tilde{\phi})
\nonumber
\\
&&-\frac{2}{16\pi}\tilde{g}^{ac}\tilde{R}_{kc}\tilde{\xi}^k+\frac{1}{16\pi}\tilde{\nabla}_b[\tilde{\nabla}^a\tilde{\xi}^b-\tilde{\nabla}^b\tilde{\xi}^a]
\end{eqnarray}
Again, using the equation of motion of the metric tensor $\tilde{g}^{bc}$ in the Einstein frame \eqref{EOM1} and contracting it with $\tilde{g}^{ab}\tilde{\xi}^c$ one finds all the other terms vanish except for the last total derivative one and, therefore, the on-shell Noether current is given by \eqref{JE1}.

\section{Derivation of the Eqs. \eqref{QAPP} and \eqref{BRAC12}} \label{APPEN2}
\subsection{Derivation of the Eq. \eqref{QAPP}}
To calculate \eqref{CHARGE} one already know that 
the Noether potential in Jordan frame is defined by \eqref{NOPJ}. If one write it in terms of the Killing and the normal vectors using $\xi^a=T\chi^a+R\rho^a$ with $R=\frac{\chi^2}{\kappa\rho^2}DT$, one obtains 
\begin{eqnarray}
&&J^{ab}=\frac{1}{16\pi}\{ \phi[\frac{2\kappa}{\chi^2}(\chi^a\rho^b-\chi^b\rho^a)T
\nonumber
\\
&&-\frac{1}{\kappa\chi^2}(\chi^a\rho^b-\chi^b\rho^a)(D^2T)] +2\xi^a\nabla^b\phi
\nonumber
\\
&&-2\xi^b\nabla^a\phi\} \label{JABVIR}
\end{eqnarray}
Using the relation $d\Sigma_{ab}=-d^2x(\chi_a\rho_b-\chi_b\rho_a)\frac{\mid \chi\mid}{\rho \chi^2}$ from (A.2) of \cite{Majhi:2011ws} one obtains
\begin{eqnarray}
&&d\Sigma_{ab}J^{ab}=-\frac{d^2x}{16\pi}\frac{\mid\chi\mid}{\rho\chi^2} [2\rho^2\phi(2\kappa T -\frac{1}{\kappa}D^2T)
\nonumber
\\
&&+(\chi_a\rho_b-\chi_b\rho_a)(2\xi^a\nabla^b\phi-2\xi^b\nabla^a\phi)] \label{DSIGJAB}
\end{eqnarray}
Using $\chi^a\nabla_a\phi=0$ and at the horizon $\rho^a\nabla_a\phi=\mathcal{O}(\chi^2)$, one gets the contributing terms at the horizon as 
\begin{align}
d\Sigma_{ab}J^{ab}=\frac{d^2x}{16\pi}2\phi(2\kappa T-\frac{1}{\kappa}D^2T) \label{SIGABJAB}
\end{align}
Therefore, one can find the desired relation \eqref{QAPP}.
\subsection{Derivation of the Eq. \eqref{BRAC12}}
As done in the previous part, writing $\xi^a$ in terms of two orthogonal vectors $\chi^a$ and $\rho^a$, one can obtain from Eq. \eqref{JAJ}
\begin{eqnarray}
&&J^a=\frac{1}{16\pi}\Big[-\frac{1}{\kappa\chi^2}(\nabla_b\phi)\chi^a\rho^b(D^2T)
\nonumber
\\
&&-\frac{2R\kappa}{\chi^2}\rho^a\rho^b
+\phi[\frac{1}{\kappa\chi^2}\rho^a(D^3T)-\frac{2\kappa}{\chi^2}\rho^a(DT)] 
\nonumber
\\
&&+2(T\chi^a+R\rho^a)\square\phi-2(\nabla^a\phi)[DT+\frac{R\kappa}{\chi^2}(\chi^2-\rho^2)]
\nonumber
\\
&&-2(T\chi^b+R\rho^b)(\nabla_b\nabla^a\phi)\Big]
\end{eqnarray}
To calculate the Lie bracket of the charges \eqref{Q1Q2}, one needs to calculate $d\Sigma_{ab}\xi^a$, which is given by
\begin{align}
d\Sigma_{ab}\xi^a=-d^2x \frac{\mid \chi\mid}{\rho\chi^2}[T\chi^2\rho_b-R\rho^2\chi_b] \label{DSIGMA}
\end{align}
Hence, a straightforward calculation gives
\begin{eqnarray}
&&d\Sigma_{ab}\xi_2^aJ_1^b=-\frac{d^2x}{16\pi} \frac{\mid \chi\mid}{\rho\chi^2}\Big\{(\nabla_a\phi)[
\frac{1}{\kappa}R_2\rho^2\rho^a(D^2T_1)
\nonumber
\\
&&-2R_1\kappa T_2\rho^2\rho^a]
+T_2\rho^2\phi[\frac{1}{\kappa}(D^3T_1)-2\kappa(DT_1)]
\nonumber
\\
&&+2(\square\phi)\rho^2\chi^2(R_1T_2-T_1R_2)
\nonumber
\\
&&-2(\nabla^b\phi)[T_2\chi^2\rho_b-R_2\rho^2\chi_b][DT_1
+\frac{R_1\kappa}{\chi^2}(\chi^2-\rho^2)]
\nonumber
\\
&&-2[T_2\chi^2\rho_b-R_2\rho^2\chi_b](T_1\chi^a+R_1\rho^a)(\nabla_a\nabla^b\phi)\Big\}
\end{eqnarray}
 Now we shall take only those terms which contribute to calculate the bracket defined in \eqref{Q1Q2} at the horizon and neglect all other terms.As $\chi_a$ is a Killing vector, $\chi^a\nabla_a\phi=0$, and at the horizon, $\rho^a\nabla_a\phi=\mathcal{O}(\chi^2)$. Therefore, some terms vanish. Taking the nonvanishing terms and using $R=\frac{\chi^2}{\kappa\rho^2}(DT)$, one obtains
\begin{eqnarray}
&&d\Sigma_{ab}\xi_2^aJ_1^b=-\frac{d^2x}{16\pi} \frac{\mid \chi\mid}{\rho\chi^2}[T_2\rho^2\phi(\frac{1}{\kappa}D^3T_1-2\kappa DT_1)
\nonumber
\\
&&+2(\square\phi)\chi^4(T_2DT_1-T_1DT_2)
\nonumber
\\
&&-2(T_2\chi^2\rho_b-\frac{\chi^2}{\kappa}(DT_2)\chi_b)(T_1\chi^a
\nonumber
\\
&&+\frac{\chi^2\rho^a}{\kappa\rho^2}(DT_1))(\nabla_a\nabla^b\phi)]
\end{eqnarray}
At the null surface $\chi^2=0$. Therefore, the second term containing $\square\phi$ does not contribute at the horizon as it is proportional to $\chi^2$. Also, the calculation gives a few symmetric terms for the interchanging of $1\leftrightarrow 2$.
At the horizon one can prove that $\chi^a\chi_b(\nabla_a\nabla^b\phi)=\rho^a\rho_b(\nabla_a\nabla^b\phi)=0$. 
Ultimately, the contributing terms to calculate the bracket \eqref{Q1Q2} is 
\begin{align}
d\Sigma_{ab}\xi_2^aJ_1^b=-\frac{d^2x}{16\pi} \frac{\mid\chi\mid\phi}{\rho}[2\kappa T_2(DT_1)-\frac{1}{\kappa}T_2(D^3T_1)]
\nonumber
\\
+\mathcal{O}(\chi^2)+(1\leftrightarrow 2 symmetric\ \ terms). \label{HBRAC}
\end{align}
So, ultimately, the bracket \eqref{Q1Q2} gives the value mentioned in \eqref{BRAC12} \\

\section{Derivation of Eq. \eqref{8}} \label{appp}
The total Lagrangian in the Jordan frame is given is given by \eqref{SJ}. We want the variation of the $\sqrt{-g}\phi R$ with respect to the first- and second-order derivative of the metric tensor $g_{ab}$, which is given by
\begin{eqnarray}
&&\delta(\sqrt{-g}\phi R)=2\phi\sqrt{-g}\delta(\partial_m\partial_ng_{pq})P^{pnmq}
\nonumber
\\
&& -2\sqrt{-g}\phi\delta(\partial_mg_{nq})[P^{nbmd}\Gamma^q_{bd}+P^{nbcd}\Gamma^m_{bc}
\nonumber
\\
&&+P^{imcq}\Gamma^n_{ic}],
\end{eqnarray}
where, $P_a^{bcd}=\frac{\partial R}{\partial R^a_{bcd}}$, and $P^{abcd}=\frac{1}{2}(g^{ac}g^{bd}-g^{ad}g^{bc})$. Now, from the straightforward calculation, one can obtain (expanding the Euler derivative)
\begin{align}
&&\partial_i[g_{ab}\frac{\delta_E \sqrt{-g}L}{\delta_E(\partial_ig_{ab})}+\partial_jg_{ab}\frac{\partial \sqrt{-g}L}{\partial(\partial_i\partial_jg_{ab})}]
\nonumber
\\
&& =\partial_i[g_{ab}\frac{\partial \sqrt{-g}L}{\partial(\partial_ig_{ab})}-g_{ab}\partial_h\frac{\partial \sqrt{-g}L}{\partial(\partial_h\partial_ig_{ab})}
\nonumber
\\
&& +\partial_jg_{ab}\frac{\partial \sqrt{-g}L}{\partial(\partial_i\partial_jg_{ab})}]=\frac{3}{16\pi}\sqrt{-g}\square\phi \label{C2}
\end{align}
Now, the surface term at the thermodynamic level is 
\begin{align}
\sqrt{-g}L_{sur}=\frac{\sqrt{-g}}{16\pi}(2\phi R^0_0-2\nabla^0\nabla_0\phi+2\square\phi)
\nonumber
\\
=\frac{1}{16\pi}\Big(2\partial_a(\phi\sqrt{-g}g^{0i}\Gamma^a_{i0})+2\sqrt{-g}\square\phi\Big)
\end{align}
From straightforward calculation, one obtains
\begin{align}
&&\partial_i[g_{ab}\frac{\delta_E \sqrt{-g}L_{sur}}{\delta_E(\partial_ig_{ab})}+\partial_jg_{ab}\frac{\partial \sqrt{-g}L_{sur}}{\partial(\partial_i\partial_jg_{ab})}]
\nonumber
\\
&& =\partial_i[g_{ab}\frac{\partial \sqrt{-g}L_{sur}}{\partial(\partial_ig_{ab})}-g_{ab}\partial_h\frac{\partial \sqrt{-g}L_{sur}}{\partial(\partial_h\partial_ig_{ab})}
\nonumber
\\
&&+\partial_jg_{ab}\frac{\partial \sqrt{-g}L_{sur}}{\partial(\partial_i\partial_jg_{ab})}] =L_{sur} \label{C4}
\end{align}
Subtracting \eqref{C4} from \eqref{C2}, one gets the mentioned result in \eqref{8}.


\begin{thebibliography}{99}
\bibitem{Riess:2001gk} 
  A.~G.~Riess {\it et al.} [Supernova Search Team Collaboration],
  ``The farthest known supernova: support for an accelerating universe and a glimpse of the epoch of deceleration,''
  Astrophys.\ J.\  {\bf 560}, 49 (2001)
  [astro-ph/0104455].
 
\bibitem{Riess:2004nr} 
  A.~G.~Riess {\it et al.} [Supernova Search Team Collaboration],
  ``Type Ia supernova discoveries at z $>$ 1 from the Hubble Space Telescope: Evidence for past deceleration and constraints on dark energy evolution,''
  Astrophys.\ J.\  {\bf 607}, 665 (2004)
  [astro-ph/0402512].
 
 \bibitem{KNOP}
 R.~A.~Knop et al.,
 ``New Constraints on $\Omega_M$, $\Omega_{\Lambda}$, and $w$ from an Independent Set of 11 High-Redshift Supernovae Observed with the Hubble Space Telescope",
  The Astrophysical Journal, 2003, {\bf 598}, 102-137.
 
\bibitem{Perlmutter:1998np} 
  S.~Perlmutter {\it et al.} [Supernova Cosmology Project Collaboration],
  ``Measurements of Omega and Lambda from 42 high redshift supernovae,''
  Astrophys.\ J.\  {\bf 517}, 565 (1999),
  [astro-ph/9812133].
  
  
\bibitem{Tonry:2003zg} 
  J.~L.~Tonry {\it et al.} [Supernova Search Team Collaboration],
  ``Cosmological results from high-z supernovae,''
  Astrophys.\ J.\  {\bf 594}, 1 (2003),
  [astro-ph/0305008].

\bibitem{Barris:2003dq} 
  B.~J.~Barris {\it et al.},
  ``23 High redshift supernovae from the IFA Deep Survey: Doubling the SN sample at $z > 0.7$,''
  Astrophys.\ J.\  {\bf 602}, 571 (2004),
  [astro-ph/0310843].
  
  
\bibitem{Perlmutter:1997zf} 
  S.~Perlmutter {\it et al.} [Supernova Cosmology Project Collaboration],
  ``Discovery of a supernova explosion at half the age of the Universe and its cosmological implications,''
  Nature {\bf 391}, 51 (1998)
  [astro-ph/9712212].

\bibitem{Riess:1998cb} 
  A.~G.~Riess {\it et al.} [Supernova Search Team Collaboration],
  ``Observational evidence from supernovae for an accelerating universe and a cosmological constant,''
  Astron.\ J.\  {\bf 116}, 1009 (1998)
  [astro-ph/9805201].

\bibitem{RIESS4}
Adam~G.~Riess et al.,
``Is there an Indication of Evolution of Type Ia Supernovaefrom their Rise Times?",
The Astronomical Journal, 1999, {\bf 118}, 2668-2674.

\bibitem{HEHL}
Friedrich~W.~Hehl, Paul~von~der~Heyde, G.~David~Kerlick, and James~M.~Nester,
``General relativity with spin and torsion: Foundations and prospects",
Reviews of Modern Physics, 1976, {\bf 48}, 393-416.

\bibitem{WILL}
C.~M.~Will,
 ``Theory and Experiment in Gravitational Physics",
 Cambridge University Press: Cambridge, UK, 1981.
 
\bibitem{Clifton:2006jh} 
  T.~Clifton,
  ``Alternative theories of gravity,''
  gr-qc/0610071.

\bibitem{Faraoni:2010yi} 
  V.~Faraoni,
  ``Black hole entropy in scalar-tensor and f(R) gravity: An Overview,''
  Entropy {\bf 12}, 1246 (2010)
  [arXiv:1005.2327 [gr-qc]].
  
\bibitem{Faraoni:1998qx} 
  V.~Faraoni, E.~Gunzig and P.~Nardone,
  ``Conformal transformations in classical gravitational theories and in cosmology,''
  Fund.\ Cosmic Phys.\  {\bf 20}, 121 (1999)
  [gr-qc/9811047].
  

  
\bibitem{Brans:1961sx} 
  C.~Brans and R.~H.~Dicke,
  ``Mach's principle and a relativistic theory of gravitation,''
  Phys.\ Rev.\  {\bf 124}, 925 (1961).
  
\bibitem{Callan:1985ia} 
  C.~G.~Callan, Jr., E.~J.~Martinec, M.~J.~Perry and D.~Friedan,
  ``Strings in Background Fields,''
  Nucl.\ Phys.\ B {\bf 262}, 593 (1985).
  
\bibitem{La:1989za} 
  D.~La and P.~J.~Steinhardt,
  ``Extended Inflationary Cosmology,''
  Phys.\ Rev.\ Lett.\  {\bf 62}, 376 (1989)
  Erratum: [Phys.\ Rev.\ Lett.\  {\bf 62}, 1066 (1989)].
  
\bibitem{Laycock:1993bc} 
  A.~M.~Laycock and A.~R.~Liddle,
  ``Extended inflation with a curvature coupled inflaton,''
  Phys.\ Rev.\ D {\bf 49}, 1827 (1994)
  [astro-ph/9306030].
  
\bibitem{Bertolami:1999dp} 
  O.~Bertolami and P.~J.~Martins,
  ``Nonminimal coupling and quintessence,''
  Phys.\ Rev.\ D {\bf 61}, 064007 (2000)
  [gr-qc/9910056].
  
  
\bibitem{Weinberg}
S.~Weinberg,
``Gravitation and Cosmology",
(Wiley, New York, 1972).  
  
\bibitem{Bhadra:2002qk} 
  A.~Bhadra,
  ``General relativity limit of the scalar tensor theories for traceless matter field,''
  gr-qc/0204014.
  
\bibitem{Faraoni:1999yp} 
  V.~Faraoni,
  ``Illusions of general relativity in Brans-Dicke gravity,
  Phys.\ Rev.\ D {\bf 59}, 084021 (1999)
  [gr-qc/9902083].
  
\bibitem{Pal:2016hxt} 
  S.~Pal,
  ``Quantized Brans-Dicke theory: Phase transition, strong coupling limit, and general relativity,''
  Phys.\ Rev.\ D {\bf 94}, no. 8, 084023 (2016)
  [arXiv:1608.06946 [gr-qc]].
  
   \bibitem{Koga:1998un} 
  J.~i.~Koga and K.~i.~Maeda,
  ``Equivalence of black hole thermodynamics between a generalized theory of gravity and the Einstein theory,''
  Phys.\ Rev.\ D {\bf 58}, 064020 (1998)
  [gr-qc/9803086].
  
\bibitem{Faraoni:2016ozb} 
  V.~Faraoni, F.~Hammad and S.~D.~Belknap-Keet,
  ``Revisiting the Brans solutions of scalar-tensor gravity,''
  Phys.\ Rev.\ D {\bf 94}, no. 10, 104019 (2016)
  [arXiv:1609.02783 [gr-qc]].
  
\bibitem{Hawking:1972qk} 
  S.~W.~Hawking,
  ``Black holes in the Brans-Dicke theory of gravitation,''
  Commun.\ Math.\ Phys.\  {\bf 25}, 167 (1972).
  
\bibitem{Santiago:1999by} 
  D.~I.~Santiago and A.~S.~Silbergleit,
  ``On the energy momentum tensor of the scalar field in scalar tensor theories of gravity,''
  Gen.\ Rel.\ Grav.\  {\bf 32}, 565 (2000)
  [gr-qc/9904003].
  
\bibitem{Dehghani:2006xt} 
  M.~H.~Dehghani, J.~Pakravan and S.~H.~Hendi,
  ``Thermodynamics of charged rotating black branes in Brans-Dicke theory with quadratic scalar field potential,''
  Phys.\ Rev.\ D {\bf 74}, 104014 (2006)
  [hep-th/0608197].

\bibitem{Sheykhi:2009vc} 
  A.~Sheykhi and M.~M.~Yazdanpanah,
  ``Thermodynamics of charged Brans-Dicke AdS black holes,''
  Phys.\ Lett.\ B {\bf 679}, 311 (2009)
  [arXiv:0904.1777 [hep-th]].
  
\bibitem{Capozziello:2010sc} 
  S.~Capozziello, P.~Martin-Moruno and C.~Rubano,
  ``Physical non-equivalence of the Jordan and Einstein frames,''
  Phys.\ Lett.\ B {\bf 689}, 117 (2010)
  [arXiv:1003.5394 [gr-qc]].
  
    \bibitem{Kang:1996rj} 
  G.~Kang,
  ``On black hole area in Brans-Dicke theory,''
  Phys.\ Rev.\ D {\bf 54}, 7483 (1996)
  [gr-qc/9606020].
  
\bibitem{Campanelli:1993sm} 
  M.~Campanelli and C.~O.~Lousto,
  ``Are black holes in Brans-Dicke theory precisely the same as a general relativity?,''
  Int.\ J.\ Mod.\ Phys.\ D {\bf 2}, 451 (1993)
  [gr-qc/9301013].

  
\bibitem{Lobo:2016izs} 
  I.~P.~Lobo,
  ``Frame transformations in Brans-Dicke theory from the viewpoint of Weyl geometry,''
  arXiv:1610.05004 [gr-qc].
  
\bibitem{Banerjee:2016lco} 
  N.~Banerjee and B.~Majumder,
  ``A question mark on the equivalence of Einstein and Jordan frames,''
  Phys.\ Lett.\ B {\bf 754}, 129 (2016)
  [arXiv:1601.06152 [gr-qc]].
  
\bibitem{Pandey:2016unk} 
  S.~Pandey and N.~Banerjee,
  ``Equivalence of Jordan and Einstein frames at the quantum level,''
  arXiv:1610.00584 [gr-qc].

\bibitem{Saltas:2010ga} 
  I.~D.~Saltas and M.~Hindmarsh,
  ``The dynamical equivalence of modified gravity revisited,''
  Class.\ Quant.\ Grav.\  {\bf 28}, 035002 (2011)
  [arXiv:1002.1710 [gr-qc]].
  
\bibitem{New} 
  S.~Bahamonde, S.~D.~Odintsov, V.~K.~Oikonomou and P.~V.~Tretyakov,
  ``Deceleration versus Acceleration Universe in Different Frames of $F(R)$ Gravity,''
  arXiv:1701.02381 [gr-qc].
  
\bibitem{Padmanabhan:2004fq} 
  T.~Padmanabhan,
  ``Holographic gravity and the surface term in the Einstein-Hilbert action,''
  Braz.\ J.\ Phys.\  {\bf 35}, 362 (2005)
  [gr-qc/0412068].
  
  \bibitem{Padmanabhan:2002jr} 
  T.~Padmanabhan,
  ``The Holography of gravity encoded in a relation between entropy, horizon area and action for gravity,''
  Gen.\ Rel.\ Grav.\  {\bf 34}, 2029 (2002)
  [gr-qc/0205090].
  
\bibitem{Mukhopadhyay:2006vu} 
  A.~Mukhopadhyay and T.~Padmanabhan,
  ``Holography of gravitational action functionals,''
  Phys.\ Rev.\ D {\bf 74}, 124023 (2006)
  [hep-th/0608120].
  
\bibitem{Bekenstein:1973ur} 
  J.~D.~Bekenstein,
  ``Black holes and entropy,''
  Phys.\ Rev.\ D {\bf 7}, 2333 (1973).
  
\bibitem{Bhattacharya:2016kbm} 
  K.~Bhattacharya and B.~R.~Majhi,
  ``Temperature and thermodynamic structure of Einstein's equations for a cosmological black hole,''
  Phys.\ Rev.\ D {\bf 94}, no. 2, 024033 (2016),
  [arXiv:1602.07879 [gr-qc]].

  
\bibitem{Wald:1993nt} 
  R.~M.~Wald,
  ``Black hole entropy is the Noether charge,''
  Phys.\ Rev.\ D {\bf 48}, no. 8, R3427 (1993)
  [gr-qc/9307038].
  
\bibitem{Brown:1986nw} 
  J.~D.~Brown and M.~Henneaux,
  ``Central Charges in the Canonical Realization of Asymptotic Symmetries: An Example from Three-Dimensional Gravity,''
  Commun.\ Math.\ Phys.\  {\bf 104}, 207 (1986).

    \bibitem{Carlip:1999cy} 
  S.~Carlip,
  ``Entropy from conformal field theory at Killing horizons,''
  Class.\ Quant.\ Grav.\  {\bf 16}, 3327 (1999)
  [gr-qc/9906126].

  
  
  
  
\bibitem{Majhi:2011ws} 
  B.~R.~Majhi and T.~Padmanabhan,
 ``Noether Current, Horizon Virasoro Algebra and Entropy,''
  Phys.\ Rev.\ D {\bf 85}, 084040 (2012)
  [arXiv:1111.1809 [gr-qc]].
  
\bibitem{Majhi:2012tf} 
  B.~R.~Majhi and T.~Padmanabhan,
  ``Noether current from the surface term of gravitational action, Virasoro algebra and horizon entropy,''
  Phys.\ Rev.\ D {\bf 86}, 101501 (2012)
  [arXiv:1204.1422 [gr-qc]].

\bibitem{Majhi:2012nq} 
  B.~R.~Majhi,
  ``Noether current of the surface term of Einstein-Hilbert action, Virasoro algebra and entropy,''
  Adv.\ High Energy Phys.\  {\bf 2013}, 386342 (2013)
  [arXiv:1210.6736 [gr-qc]].
  
\bibitem{Majhi:2013lba} 
  B.~R.~Majhi and S.~Chakraborty,
  ``Anomalous effective action, Noether current, Virasoro algebra and Horizon entropy,''
  Eur.\ Phys.\ J.\ C {\bf 74}, 2867 (2014)
  [arXiv:1311.1324 [gr-qc]].
  
\bibitem{Majhi:2014lka} 
  B.~R.~Majhi,
  ``Conformal Transformation, Near Horizon Symmetry, Virasoro Algebra and Entropy,''
  Phys.\ Rev.\ D {\bf 90}, no. 4, 044020 (2014)
  [arXiv:1404.6930 [gr-qc]].

\bibitem{Majhi:2015tpa} 
  B.~R.~Majhi,
  ``Near horizon hidden symmetry and entropy of Sultana-Dyer black hole: A time dependent case,''
  Phys.\ Rev.\ D {\bf 92}, no. 6, 064026 (2015)
  [arXiv:1505.03310 [gr-qc]].
  
\bibitem{Majhi:2017fua} 
  B.~R.~Majhi,
  ``Noncomutativity in near horizon symmetries in gravity,''
  Phys.\ Rev.\ D {\bf 95}, no. 4, 044020 (2017)
  [arXiv:1701.07952 [gr-qc]].

\bibitem{Carlip:1998qw} 
  S.~Carlip,
  ``What we don't know about BTZ black hole entropy,''
  Class.\ Quant.\ Grav.\  {\bf 15}, 3609 (1998)
  [hep-th/9806026].
  
\bibitem{Kolekar:2010dm} 
  S.~Kolekar and T.~Padmanabhan,
  ``Holography in Action,''
  Phys.\ Rev.\ D {\bf 82}, 024036 (2010)
  [arXiv:1005.0619 [gr-qc]].


  
\bibitem{Padmanabhan:2002sha} 
  T.~Padmanabhan,
  ``Classical and quantum thermodynamics of horizons in spherically symmetric space-times,''
  Class.\ Quant.\ Grav.\  {\bf 19}, 5387 (2002)
  [gr-qc/0204019].
  
\bibitem{Jacobson:1993pf} 
  T.~Jacobson and G.~Kang,
  ``Conformal invariance of black hole temperature,''
  Class.\ Quant.\ Grav.\  {\bf 10}, L201 (1993)
  [gr-qc/9307002].

\bibitem{Brown:1992br} 
  J.~D.~Brown and J.~W.~York, Jr.,
  ``Quasilocal energy and conserved charges derived from the gravitational action,''
  Phys.\ Rev.\ D {\bf 47}, 1407 (1993)
  [gr-qc/9209012].
  
\bibitem{Faraoni:2014lsa} 
  V.~Faraoni and V.~Vitagliano,
  ``Horizon thermodynamics and spacetime mappings,''
  Phys.\ Rev.\ D {\bf 89}, no. 6, 064015 (2014)
  [arXiv:1401.1189 [gr-qc]].
  
\bibitem{Prain:2015tda} 
  A.~Prain, V.~Vitagliano, V.~Faraoni and M.~Lapierre-Léonard,
  ``Hawking–Hayward quasi-local energy under conformal transformations,''
  Class.\ Quant.\ Grav.\  {\bf 33}, no. 14, 145008 (2016)
  [arXiv:1501.02977 [gr-qc]].
  
\bibitem{Faraoni:2015sja} 
  V.~Faraoni,
  ``Quasilocal energy in modified gravity,''
  Class.\ Quant.\ Grav.\  {\bf 33}, no. 1, 015007 (2016)
  [arXiv:1508.06849 [gr-qc]].
    
\bibitem{Hammad:2016yjq} 
  F.~Hammad,
  ``More on the conformal mapping of quasi-local masses: The Hawking-Hayward case,''
  Class.\ Quant.\ Grav.\  {\bf 33}, no. 23, 235016 (2016)
  [arXiv:1611.03484 [gr-qc]].
  
\bibitem{Bose:1998yp} 
  S.~Bose and D.~Lohiya,
  ``Behavior of quasilocal mass under conformal transformations,''
  Phys.\ Rev.\ D {\bf 59}, 044019 (1999)
  [gr-qc/9810033].
\\
\\
  \\

















  
  

  
  
 
  
  





  

  

  

  
  
\end{thebibliography}
\end{document}